\documentclass[prfluids,showpacs,notitlepage,byrevtex,longbibliography]{revtex4-1}
\usepackage{epsf,graphicx,amsmath,amssymb}
\usepackage[usenames]{color}
\usepackage[utf8]{inputenc}  
\usepackage[T1]{fontenc}
\DeclareMathOperator{\arcsech}{arcsech} 

\newcommand{\sech}{\operatorname{sech}}

\begin{document}

\title{Coexistence of solitons and extreme events in deep water surface waves}
%bien: Coexistence of solitons and extreme events in deep water surface waves ///(random gravity)
% Papier court: Experimental observation of integrable "turbulence" in deep water surface waves 
%non: Observation of soliton turbulence in deep water surface waves

%\author{A. Cazaubiel$^1$, G. Michel$^2$, S. Lepot$^3$, B. Semin$^2$, S. Auma{\^i}tre$^3$,\break
%M. Berhanu$^1$, P. Suret$^4$, S. Randoux$^4$, F. Bonnefoy$^5$ and E. Falcon$^1$\email[E-mail: ]{eric.falcon@univ-paris-diderot.fr}}
%
%\affiliation{$^1$Univ Paris Diderot, Sorbonne Paris Cit\'e, MSC, UMR 7057 CNRS, F-75 013 Paris, France\\
%$^2$Ecole Normale Sup\'erieure, LPS, UMR 8550 CNRS, F-75 205 Paris, France\\
%$^3$CEA-Saclay, Sphynx, DSM, URA 2464 CNRS, F-91 191 Gif-sur-Yvette, France \\
%$^4$Université de Lille, PHLAM, UMR 8523 CNRS, F-59655 Villeneuve d’Ascq, France\\
%$^5$Ecole Centrale de Nantes, LHEEA, UMR 6598 CNRS, F-44 321 Nantes, France
%}

\author{A. Cazaubiel$^1$, G. Michel$^2$, S. Lepot$^3$, B. Semin$^2$, \break
S. Auma{\^i}tre$^3$, M. Berhanu$^1$, F. Bonnefoy$^4$, and E. Falcon$^1$}
\email{eric.falcon@univ-paris-diderot.fr}

\affiliation{$^1$Univ Paris Diderot, Sorbonne Paris Cit\'e, MSC, UMR 7057 CNRS, F-75 013 Paris, France\\
$^2$Ecole Normale Sup\'erieure, LPS, UMR 8550 CNRS, F-75 205 Paris, France\\
$^3$CEA-Saclay, Sphynx, DSM, URA 2464 CNRS, F-91 191 Gif-sur-Yvette, France \\
$^4$Ecole Centrale de Nantes, LHEEA, UMR 6598 CNRS, F-44 321 Nantes, France
}
\date{\today}

\begin{abstract}
We study experimentally, in a large-scale basin, the propagation of unidirectional deep water gravity waves stochastically modulated in phase. We observe the emergence of nonlinear localized structures that evolve on a stochastic wave background. Such a coexistence is expected by the integrable turbulence theory for the nonlinear Schrödinger equation (NLSE), and we report the first experimental observation in the context of hydrodynamic waves. We characterize the formation, the properties and the dynamics of these nonlinear coherent structures (solitons and extreme events) within the incoherent wave background. The extreme events result from the strong steepening of wave train fronts, and their emergence occurs after roughly one nonlinear length scale of propagation (estimated from NLSE). Solitons arise when nonlinearity and dispersion are weak, and of the same order of magnitude as expected from NLSE. We characterize the statistical properties of this state. The number of solitons and extreme events is found to increase all along the propagation, the wave-field distribution has a heavy tail, and the surface elevation spectrum is found to scale as a frequency power-law with an exponent $-4.5\pm 0.5$. Most of these observations are compatible with the integrable turbulence theory for NLSE although some deviations (e.g. power-law spectrum, asymmetrical extreme events) result from effects proper to hydrodynamic waves.
\end{abstract}
\maketitle
%\pacs{45.70.-n, 81.05.Rm, 05.20.Dd, 75.50.-y{\color{red} to be changed}}
%Sujet: L4-45: Fluid Dynamics
%PhysSH: Fluid Dynamics - Turbulence -Internal flows

 %occurs from the initial incoherent waves. 
%We report on an experimental observation of a new statistical state for the propagation of unidirectional gravity waves in a deep water regime where extreme events and enveloppe solitons coexist within a sea of stochastic waves. Such a state expected theoretically by integrable turbulence theory for Non Linear Schrödinger Equation (NLSE) has never been observed experimentally so far for hydrodynamics waves in a deep water regime. In our large-scale experiment, we can control the nonlinearity-to-dispersion ratio in order to observe solitons governed by NLSE. 
 
\section{Introduction}
When many random and weakly nonlinear dispersive waves propagate and interact with one another, several statistically stationary states are theoretically predicted such as weak wave turbulence, statistical equilibrium, or integrable turbulence. Weak wave turbulence describes an ensemble of nonlinear waves undergoing resonant interactions. These energy transfers between spatial and temporal scales lead generally to a cascade of wave energy from a large (forcing) scale, up to a small (eventually dissipative) one. This phenomenon occurs in various situations ranging from spin waves in solids, internal or surface waves in oceanography up to plasma waves in astrophysics (for reviews, see~\cite{Falcon2010,Zakharovbook,Nazarenkobook,Newell2011}). The theory of weak wave turbulence, developed in the 1960s \cite{Hasselmann1962,Benney1967,Zakharov1967}, leads to analytical predictions on the wave energy spectrum in a stationary state, and has since been applied in almost all domains of physics involving waves \cite{Zakharovbook,Nazarenkobook}. In the past decade, an important experimental effort has been devoted to test the domain of validity of weak turbulence theory on different wave systems (e.g. hydrodynamics, optics, hydro-elastic or elastic waves) \cite{NazarenkoAdvance2013}. In the absence of an inverse cascade, it also predicts the equipartition of energy at scales larger than the forcing one, which has been recently observed experimentally \cite{Michel17}.

The theory of integrable turbulence combines the above statistical approach together with the property of integrability of an equation showing soliton solutions [e.g. Korteweg-de Vries equation (KdVE) or nonlinear Schrödinger equation (NLSE)] \cite{Zakharov71,Zakharov09}. Even though no dissipation or forcing term is part of such equations, random initial conditions generally do not relax toward thermal equilibrium \cite{Randoux16}. Instead, the emergence and the dynamics of a large number of nonlinear coherent structures (such as solitons or breathers) from the incoherent waves forms a statistical state, called integrable turbulence. It has been encountered in various situations ranging from plasma waves \cite{Kingsep73} to optical waves \cite{Schwache97,Randoux14,Toenger15,Walczak15,Suret16}. This state is different from the wave turbulence one, since no resonant wave interaction occurs, and no constant flux of a conserved quantity cascades through the scales \cite{Zakharov09}.

%By injecting random initial conditions to these equations, the wave spectrum generally relaxes to a statistically stationary state different from the standard thermal equilibrium or from the out-of-equilibrium one of wave turbulence 
 
In the context of surface waves on a fluid, KdVE describes unidirectional long waves in shallow water, whereas NLSE describes unidirectional nonlinear wave packets of arbitrary depth (although the nature of the solutions in the shallow and deep water limits strongly differs, see e.g. \cite{Chabchoub2013}). In the shallow water regime, beyond numerical confirmations \cite{Osborne93}, direct experimental verifications of integrable turbulence have been performed recently in field experiments \cite{Osborne91,Costa14} and in the laboratory \cite{Perrard15,Hassaini17}. For deep water gravity waves, the development of modulational instability is predicted to generate a state intermediate between weak turbulence and the superposition of weakly interacting solitons \cite{Zakharov09}. The statistic of such state have been the subject of several experiments, in which the waves are forced with noise. %In deep water, several experiments were performed to find the statistics of unidirectional nonlinear waves forced with non deterministic initial conditions. %one-dimensional wave turbulence is predicted to be unstable and should coexist with solitons \cite{Zakharov04},
Non Gaussian statistics of the wave height were observed to emerge from such a random forcing \cite{Onorato04,Shemer09,ShemerPoF10,ShemerJGR10,ElKoussaifiPRE18} as predicted theoretically from NLSE with random initial conditions \cite{JanssenJPO03}. Direct numerical simulations of random waves with NLSE have been also reported \cite{Onorato01,Dysthe03}. Time series from field measurements in the ocean were compared to NLSE to search for solitons and their possible link with extreme wave appearance \cite{Slunyaev06}. The highest waves that may appear in a chaotic wave field, called rogue waves, are indeed a question of intense debate \cite{SotoCrespo16,Onorato01,Slunyaev06,Onorato04,Osborne05,Islas05}. Although all of these experimental studies were not explicitly compared against the predictions of integrable turbulence, NLSE roughly captures the reported wave statistics of unidirectional random waves. However, neither the identification of the coherent structures involved in the integrable turbulence regime, nor the deviation of real systems from these predictions, have been experimentally studied. Such deviations should be more easily highlighted in hydrodynamics than in nonlinear optics, since more approximations are needed to reduce the dynamics to NLSE.
%\textbf{This second point is crucial in hydrodynamics, since the reduction of the dynamics to NLSE requires more approximations than in nonlinear optics. [Reformuler ? ]}

Here, we study experimentally the propagation of an unidirectional deep water carrier wave stochastically modulated in phase. Waves of narrow spectral bandwidth are generated at one end of the tank, and the evolution of the statistical properties of the wave field is measured along the propagation. The range of parameters and the design of the experiment are set to observe solitons governed by NLSE, the linear (dispersive) and nonlinear time scales of propagation being controllable experimentally and chosen of the same order of magnitude.  We show that a spontaneous formation of coherent localized structures, such as solitons and extreme events, occurs from the initial incoherent waves. We characterize the emergence, the property and the dynamics of these solitons and extreme events immersed in a sea of smaller stochastic waves. Such a coexistence between erratic waves and coherent structures is expected from NLSE integrable turbulence \cite{Zakharov09}, and has been reported in optics \cite{Suret16}. After one and a half nonlinear length scale of propagation, we observe a heavy-tailed (distance independent) distribution of the wave field statistics, as expected by integrable turbulence, and the emergence of extreme events. The experimental wave spectrum is then found to follow a frequency power-law with an exponent $-4.5\pm 0.5$, and we show that this feature traces back to the strong steepening of the waves. This power law spectrum, as well as the occurrence of highly asymmetrical extreme events, are not described by NLSE but comes from the specific features of hydrodynamics waves. For instance, the spectrum exponent is probably related to the random modulation of the harmonics (bound waves) of the wave field. 

%{\color{blue}This} transfer mechanism has been reported theoretically \cite{Zakharov04,Rumpf09,Newell12}, and differs from wave interactions occurring in wave turbulence.

The article is organized as follows. We first recall some theoretical results of 1D NLSE for deep water waves. Then, we give estimates of the typical propagation time scales of the problem. We describe the experimental setup, then the experimental results, before discussing our results with respect to the integrable turbulence theory. 

%The study of the spatial evolution of a Gaussian-shaped wave packet with random phases \cite{Shemer09,Shemer10,Shemer10b}
 
\section{1D nonlinear Schrödinger equation for deep water waves}
In a deep water regime ($kh \gg 1$), the dispersion relation of linear gravity waves reads 
\begin{equation}
\omega_{lin}(k)=\sqrt{gk} {\rm \ ,}
\label{displin}
\end{equation}
with the fluid depth, $h$, the acceleration of gravity, $g$, the wavenumber, $k=2\pi/\lambda$, the angular frequency, $\omega=2\pi f$, the frequency, $f$, and wavelength, $\lambda$, of the wave.

Assume a linear monochromatic wave of wavenumber $k_0$, and angular frequency $\omega_0\equiv \omega_{lin}(k_0)$. Its phase velocity, $c=\omega_0/k_0=\sqrt{g/k_0}$ thus increases as the square root of its wavelength, the group velocity being $c_g=d\omega_0/dk_0=c/2$. When the wave amplitude $a$ is not much smaller than $\lambda$, nonlinear terms in the Euler equations have to be taken into account. The dispersion relation of 
a progressive periodic wave (the so-called Stokes wave), then reads \cite{Whitham}
\begin{equation}
\omega(k)=\omega_{lin}(k) \left[1+\frac{k^2a^2}{2}+ O(k^4a^4)\right]. \label{Stokes} 
\end{equation}
 %its phase velocity $c=c_0\sqrt{1+k^2a^2/2}$, and its group velocity $C_g=c_{g}\sqrt{1+5k^2a^2/2}$. These relations depend on the wave steepness.

Consider a 1D nonlinear wave train with a complex envelope $A$ slowly varying in time $T$ and space $X$ with respect to the carrier wave ($\omega_0$, $k_0$), i.e.
\begin{equation}
\eta(x,t)=\frac{1}{2}\left [ A(X,T)e^{i(\omega_0t-k_0x)} + {\rm c.c.} \right ] {\rm \ \cdot}
\label{defEta}
\end{equation}
Here, $X=\epsilon x$ and $T=\epsilon t$ where $\epsilon \ll 1$ is a dimensionless parameter enforcing the slow space and time modulation, and c.c. denotes the complex conjugate. The small parameter $\epsilon$ is chosen to also be the steepness of the carrier, i.e. $\epsilon=k_0A$. Substituting $a$ by $A$ in the dispersion relation of the Stokes wave leads to  %Performing Taylor series expansion of the above nonlinear dispersion relation around the wavenumber $k_0$ of the carrier sinusoidal wave, and about the initial envelope amplitude $A_0=A(0,0)$,  
\begin{equation}
\omega(k)=\omega_{lin}(k)\left(1+\frac{k^2|A|^2}{2}\right) {\rm \ \cdot}
\label{NLdispersion}
\end{equation}
Now, expanding Eq.\ (\ref{NLdispersion}) into Taylor series expansion about $k_0$, and about the initial amplitude $A_0\equiv A(0,0)=0$ leads to \cite{Remoissenet99}
\begin{equation}
%\begin{multline}
\omega(k,|A|^2)-\omega_0 =\left.\frac{\partial \omega}{\partial k}\right|_{k_0}(k-k_0) + \left.\frac{1}{2}\frac{\partial^2 \omega}{\partial k^2}\right|_{k_0}(k-k_0)^2 \\ + \left.\frac{\partial \omega}{\partial |A|^2}\right|_{|A_0|^2}(|A|^2-|A_0|^2) {\rm \ \cdot}
%\end{multline}
\end{equation}
Using the notations, $\Omega=\omega-\omega_0$ and $K=k-k_0$, the dispersion relation of the modulated wave reads
\begin{equation}
\Omega(K,|A|^2)=c_gK+PK^2-Q|A|^2 {\rm \ ,}
\label{RDenvelope}
\end{equation}
valid in the vicinity of $\omega_0$ and $k_0$, with $c_g\equiv \partial \omega / \partial k|_{k=k_0}$, $P\equiv \partial^2 \omega/2\partial k^2|_{k=k_0}$, and $Q\equiv -\partial \omega/\partial |A|^2|_{A_0=0}$. All these parameters are known using the nonlinear dispersion relation of Eq.~(\ref{NLdispersion}). Following \cite{Remoissenet99}, we use the properties of the Fourier transforms for the envelope ($K=-i\epsilon \partial /\partial X$ ; $\Omega=-i\epsilon \partial /\partial T$), substitute these relationships in Eq.~(\ref{RDenvelope}), and apply the resulting operator to $A$. At order $O(\epsilon^2)$ in dispersive and nonlinear terms, the wave train envelope $A$ is then governed by NLSE \cite{Benney67,Zakharov68}
\begin{equation}
i\left(\frac{\partial A}{\partial t} + c_g\frac{\partial A}{\partial x}\right) - P\frac{\partial^2 A}{\partial x^2} - Q|A|^2A = 0 {\rm \ ,}
\label{NLS}
\end{equation}
with $c_g=\omega_0/(2k_0)$ the group velocity of the wave packet, $P=-\omega_0/(8k^2_0)$ the dispersive parameter, and $Q=-\omega_0k_0^2/2$ the nonlinear one. Note that the variables $X$ and $T$ have been put in lower case in Eq.~(\ref{NLS}) for easier reading thereafter. This equation is integrable, and an inverse scattering transform (IST) can solve Eq.~(\ref{NLS})~\cite{Zakharov68}.
 
 \begin{figure}
\includegraphics[width=7.2cm]{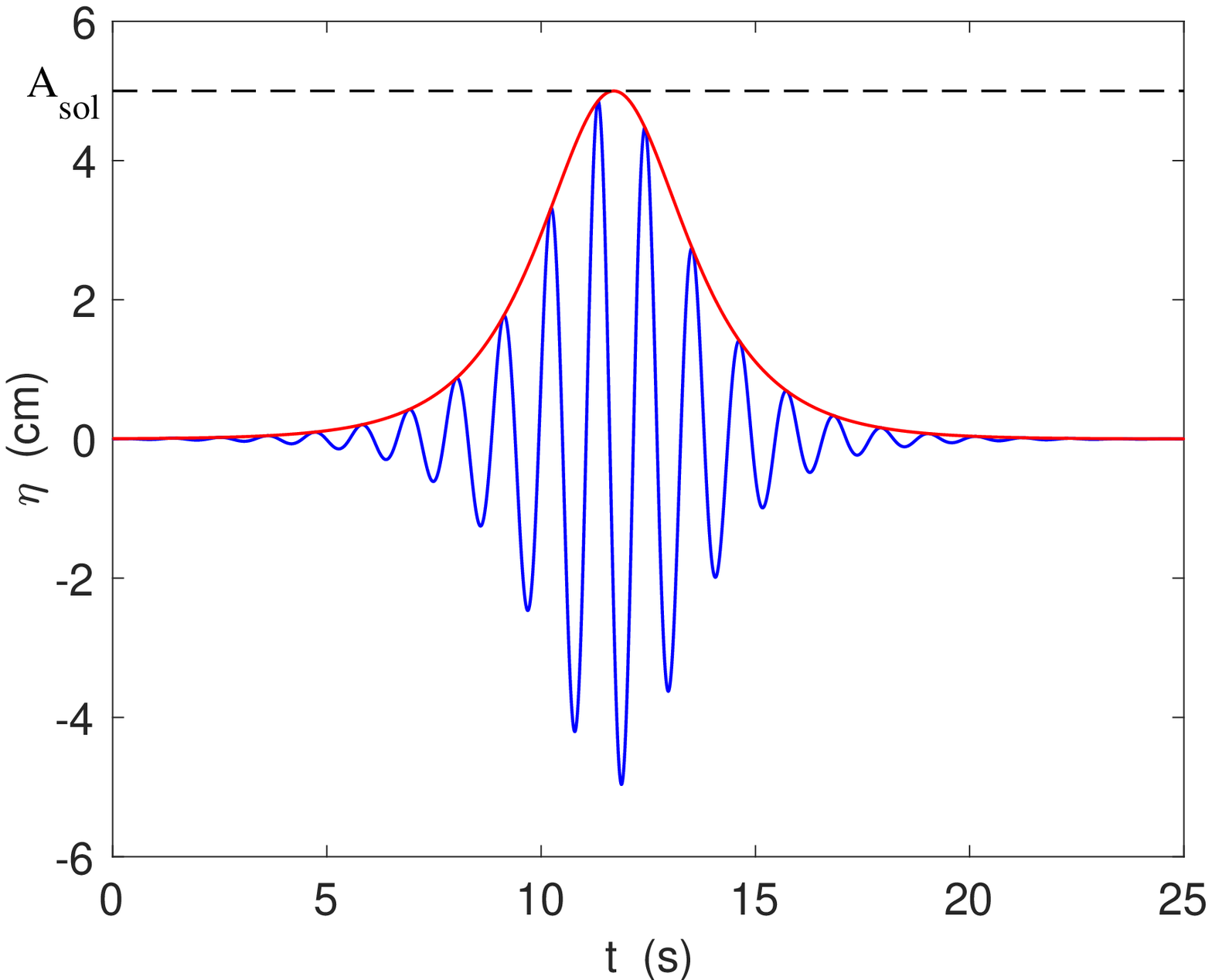}%{Fig1_left-2.eps}
\includegraphics[width=7cm]{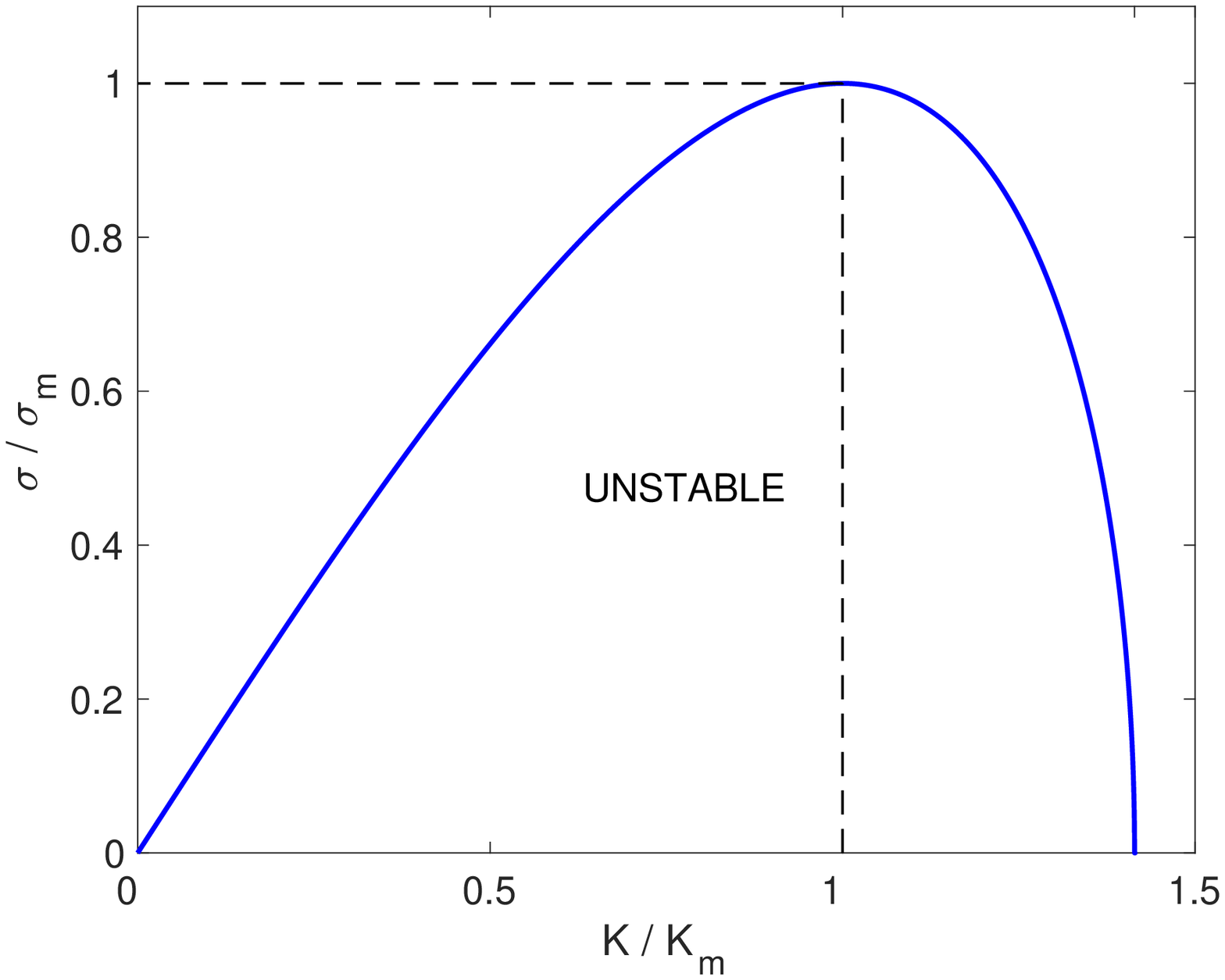}%{Fig1_right-1.eps}
\caption{Left: Theoretical envelope soliton $\eta_{{\rm sol}}(t)$ of Eq. (\ref{soliton}) with $f_0=0.9$ Hz, $x=10$ m and $A_{{\rm sol}}=5$ cm. Red solid line shows the envelope of the solution. Right: dimensionless growth rate of the Benjamin-Feir instablility, $\sigma/\sigma_m$, versus dimensionless wavenumber $K/K_m$. Unstable modes are found below the solid line.}
 \label{fig01}
 \end{figure}
 
In the deep water regime, the product of the dispersive and nonlinear term, $PQ$, is always positive. This regime, called focusing or anormal regime in optics, selects a type of solutions of Eq.\ (\ref{NLS}). The latter admits envelope soliton solution ($\sech$-shaped pulse), spatially localized, and the corresponding wave profile is of the form \cite{Remoissenet99,ZakharovShabat72}
%\begin{multline}
\begin{equation}
\eta_{{\rm sol}}(x,t)=A_{\rm sol} \sech \left[\sqrt{2}k_0^2A_{\rm sol}(x-c_gt)\right]\times \\ \cos\left[\omega_0\left(1+\frac{k_0^2A_{\rm sol}^2}{4}\right)t-k_0x\right] {\rm \ \cdot}
\label{soliton}
\end{equation}
%\end{multline}
$A_{\rm sol}$ being the maximum of the envelope soliton. Its full width at half maximum (FWHW) is then 
\begin{equation}
L_{\rm sol}=\sqrt{2}\arcsech(1/2)/(A_{\rm sol}k_0^2).%$L_{{\rm sol}}=1/[\sqrt{2}\sech(1/2)A_0k_0^2]$.
\label{Lsol} 
\end{equation}
This envelope soliton is shown in Fig.\ \ref{fig01}(left). It was first observed in deep water \cite{Yuen75} and then in nonlinear electrical transmission lines \cite{Yagi76}. Since then, others soliton solutions of the focusing NLSE, localized in both the space and time domains, have been derived \cite{ChabchoubPOF2016} (such as the Peregrine soliton \cite{Peregrine83}, Kuznetsov-Ma breathers \cite{Kuznetsov,Ma} and Akmediev breathers \cite{Akmediev1985,Akmediev1987}) and observed experimentally \cite{Chabchoub11,Slunyaev13}.  For instance, the Peregrine soliton reads \cite{Chabchoub11}
\begin{equation}
\eta_{{\rm p}}(x,t)=\Re\left\{A_{\rm p}\exp{\left(-\frac{ik_0^2A_{\rm p}^2\omega_0 t}{2}\right)}\left[ 1 - \frac{4(1-ik_0^2A_{\rm p}^2\omega_0 t)}{1+[2\sqrt{2}k_0^2A_{\rm p}(x-c_gt)]^2 + k_0^4A_{\rm p}^4\omega_0^2 t^2}  \right] \exp{\left[i(k_0x-\omega_0 t)\right]} \right\} {\rm \ ,}
\label{solitonP}
\end{equation}
$A_{\rm p}$ being the maximum of the Peregrine soliton. Its maximum amplification, which occurs at $x=0$ and $t=0$, is a factor of 3 higher than the background carrier wave. Its dynamics was first reported in nonlinear fibers \cite{KiblerNature10}, then in water wave tanks \cite{Chabchoub11} and plasmas \cite{BailungPRL11}. Moreover, Eq. (\ref{NLS}) also admits constant envelope solutions which correspond to uniform sinusoidal wave train solution of constant amplitude $a_0$ with the leading order correction to the angular frequency introduced in Eq.\ \eqref{Stokes},
\begin{equation}
\eta(x,t)=a_0\cos\left[\omega_0\left(1+\frac{k_0^2a_0^2}{2}\right)t -k_0x\right] {\rm \ ,}
\end{equation}
which may be modulationnally unstable if $\Omega^2=(K^2-2a_0^2Q/P)P^2K^2<0$, that is for $0<|K|<|K_c|\equiv a_0\sqrt{2Q/P}=2\sqrt{2}a_0k_0^2$. The maximum growth rate of the instability is achieved when $\partial\Omega^2/\partial K=0$, that is for $K_m=a_0\sqrt{Q/P}=K_c/\sqrt{2}=2a_0k_0^2$ \cite{Remoissenet99}. The growth rate $\sigma=-\Omega^2$ is maximum for $\sigma_m=a_0^2Q=\omega_0(a_0k_0)^2/2$. Fig.\ \ref{fig01}(right) shows the theoretical growth rate of the instability $\sigma/\sigma_m$ {\em vs.} $K/K_m$. In frequency space, with the use of the group velocity, the instability occurs for $0 <\Omega < \sqrt{2}\omega_0a_0k_0$ \cite{Annexe0}. The quasi-plane wave instability with respect to slowly modulating perturbation is due to the interplay between nonlinearity and dispersion. It has been first discovered by Lighthill \cite{Lighthill65}, and called modulation instability, but it is often referred to as the Benjamin-Feir instability since it was Benjamin and Feir who first applied it to surface waves in the limit of vanishing steepness ($a_0k_0\rightarrow 0$) \cite{BenjaminFeir67}. Several experiments performed in deep water have successfully verified this instability prediction \cite{BenjaminFeir67,Lake77,Melville82,LonguetHiggins80,Su82}. In the Fourier space, the modulation instability consists of a pair of sideband components growing around the angular frequency of the carrier wave $\omega_0$.  %JanssenJPO03

%\begin{figure}
%\includegraphics[width=7cm]{GrowthRate}
%\caption{Nonlinear dispersion relation of the perturbation to $A_0$. This corresponds also to the normalized growth rate of the Benjamin-Feir instablility versus normalized wave-number. Unstable modes are found below the solid line.}
%\label{fig02}
%\end{figure}

Let us now introduce the linear and nonlinear propagation time scales. Balancing the first and last terms of Eq.\ (\ref{NLS}), the nonlinear timescale reads 
\begin{equation}
T_{nl}=\frac{1}{QA_0^2}=\frac{2}{\omega_0 \epsilon^2} {\rm \ ,}
\label{Tnl}
\end{equation}
where $\epsilon \equiv k_0|A_0|$ corresponds to the steepness of the carrier. Balancing the first and third terms of Eq.\ (\ref{NLS}), the linear or dispersive timescale reads 
\begin{equation}
T_{lin}=\frac{\Delta L^2}{2P}=\frac{4\frac{k_0^2}{\Delta k^2}}{\omega_0}  {\rm \ ,}
\label{Tlin}
\end{equation}
where $\Delta L$ is the typical size of a modulation (i.e. the half-width of a Gaussian envelope at an amplitude of $|A_0|/\sqrt{e}$), $\Delta k=1/\Delta L$ thus stands for the typical spectral bandwidth of the modulation. The factor $1/2$ in Eq.\ (\ref{Tlin}) comes from the dispersion-induced spreading of a Gaussian pulse governed by Eq.\ (\ref{NLS}) with $Q=0$ \cite{Annexe}.
Using Eqs.\ (\ref{Tnl}) and (\ref{Tlin}), the ratio of both times thus reads
\begin{equation}
\frac{T_{lin}}{T_{nl}}=\frac{2\epsilon^2}{(\Delta k/k_0)^2} {\rm \ \cdot}
\label{Tratio}
\end{equation}
This ratio gives the degree of nonlinearity of the wave propagation. It is also related to the Benjamin-Feir index (BFI) of the modulation instability defined as $BFI_k\equiv 2\epsilon/(\Delta k/k_0)$ for random waves of narrow spectral bandwidth \cite{Chabchoub15}. It quantifies the ratio between the wave steepness to the normalized spectral width of the initial condition. When ${\rm BFI}_{k} > 1/\sqrt{2}$, the modulation instability at the most unstable wavenumber occurs. Indeed, in this case, one has $\Delta k < 2\sqrt{2}A_0 k_0^2$ as found above for a monochromatic wave. In the frequency space, the BFI reads $BFI_{\omega}= \epsilon/(\Delta \omega/\omega_0)$ \cite{Annexe3}, and the instability occurs for ${\rm BFI}_{\omega} > 1/\sqrt{2}$, that is $\Delta \omega < 2\omega_0A_0 k_0$. For narrow spectral bandwidth processes and from Eq.\ (\ref{defEta}), the relation between the random surface elevation, $\eta$, and its envelope is $\langle |A|^2\rangle=2 \langle \eta^2 \rangle \equiv 2\sigma_{\eta}^2$, with $\sigma_{\eta}$ the rms value of $\eta(t)$, and $\langle \cdot \rangle$ an average over time. This leads to the use of $\epsilon=\sqrt{2}\epsilon_{\eta}$ in the above definitions of BFI, as used in experiments \cite{JanssenJPO03,Onorato04}, with $\epsilon_{\eta}\equiv k_0\sigma_{\eta}$ the initial steepness.

  \begin{figure}
   \centering
  \includegraphics[width=14 cm]{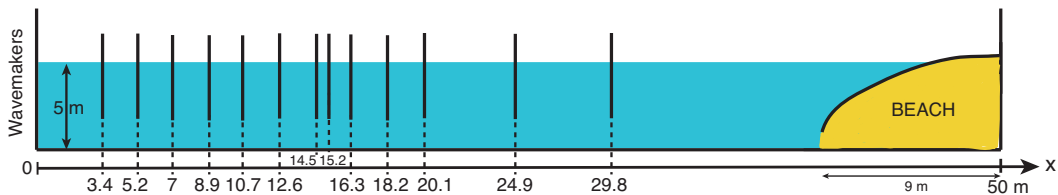}%Setup2.eps
  \caption{Sketch of a vertical section of the wave basin facility at Ecole Centrale de Nantes and locations of the resistive probes.}
  \label{DispoExp}
  \end{figure}

The characteristic length scales of the problem are related to the typical time scales by the group velocity: $L_{lin}=c_gT_{lin}=2k_0/\Delta k^2$ and $L_{nl}=c_gT_{nl}=1/(k_0\epsilon^2)$. Finally, note that the typical timescale of the carrier modulation is related to the nonlinear timescale of the problem. Indeed, one has $\Omega_m=1/T_{nl}$, with $\Omega_m$ the modulation frequency at the maximum growth rate of the modulation instability. The corresponding wavenumber $K_m$ is also related to the inverse of the width of the envelope soliton of Eq.\ (\ref{soliton}).    

%\textbf{Finally, note that the modulation frequency at which the growth rate of the instability is maximum is related to the nonlinear timescale since $\Omega_m=1/T_{nl}$. The typical width of the modulation instability and of the envelope-soliton solution of Eq.\ (\ref{soliton}) is also linked. Indeed, the width of a pulse envelope soliton of amplitude $a_0$ of Eq.\ (\ref{soliton}) corresponds approximatively to $1/K_m$. [A reformuler ?]}

%{\color{red}Finally, a dimensionless form of the NLS Eq.\ (\ref{NLS}), $i\tilde{A}_{\tilde{T}} - \tilde{A}_{\tilde{X}\tilde{X}} - {BFI_k}^2 |\tilde{A}|^2\tilde{A} = 0$, is obtained in the reference frame moving with the group velocity, and using the rescaled variables: $\tilde{A}\equiv A/\eta_{{\rm rms}}$, $\tilde{X}\equiv  x\Delta k$ and $\tilde{T}=t/(2T_{lin})$. to be checked}

\section{Experimental set up}
The experimental setup is shown in Fig. \ref{DispoExp} and is similar to the one described in Ref. \cite{BonnefoyJFM16}. Experiments are carried out in a large-scale wave basin (50 m long $\times$ 30 m wide $\times$ 5 m deep) at Ecole Centrale de Nantes. At one end of the basin is a wavemaker made of 48 independently controlled flaps, %that are hinged 2.8 m below the free surface, 
whereas an absorbing sloping beach strongly reduces wave reflections at the opposite end. We mechanically generate a 1D monochromatic carrier wave randomly modulated in phase and in amplitude. The carrier frequency is set to $f_0=0.9$ Hz corresponding, using Eq.\ (\ref{displin}), to a wavelength of $\lambda_0=1.9$ m in a deep water regime ($k_0 h \approx 16$). The period of the carrier, $T_0=1/f_0$, is 1.1 s. The modulation of this carrier is  slow compared to $T_0$, i.e. of narrow frequency spectral bandwidth $\Delta f$ (with $\Delta f / f_0 < 0.26$). %$\sigma_f$ ($\sigma_f/f_0 \leq 0.1$).
More precisely, the wavemaker is driven to reproduce the wave profile $\eta(x=0,t)=\eta_0(t)$ in front of it ($x=0$) with the prescribed wave steepness using the Fourier modes
\begin{equation}
\eta_0(t)=\sum_{n=1}^{N} \, a_{n} \, \cos  (2\pi f_n t+\phi_n)
\label{Forcing1}
\end{equation}
where $f_n=n/ \theta$ is  the frequency of the $n$th Fourier mode and $\phi_n$ is a phase chosen randomly from a uniform distribution in the interval $[0;2\pi]$. The fundamental period of the Fourier series is $\theta=2048$ s and $N$ is the number of wave components. The Fourier mode amplitude spectrum $a_{n}$ is chosen to be a narrow-banded Gaussian spectrum centered on  $f_0$ given by
\begin{equation}
a_{n}=A \, \exp \left[ -\frac{1}{2} \left(\frac{f_n-f_0}{\Delta f/(2\sqrt{2\ln{2}})} \right)^{2}\right]
\label{Forcing2}
\end{equation}
where $\Delta f$ is the full width of this spectrum at half maximum and $A$ a scale factor. The latter is adjusted so the standard deviation wave elevation $\eta_0(t)$ is $\epsilon_{\eta}/(\sqrt{2} k_0)$.

%$$
%\eta(t,x=0)=\sum_{n=1}^{2N}a_{0_{n}}\cos \left( 2\pi nf_0t/N + \phi_n\right)
%$$
%where $\phi_n$ is the phase of the $n$th Fourier mode chosen randomly from a uniform distribution in the interval $[0;2\pi)$, and $a_{0_{n}}$ the corresponding Fourier mode amplitudes that are random with average values related to the prescribed wave steepness. The generated wave field has a narrow-banded Gaussian spectrum centered on $f_0$. %and of standard deviation $\Delta f/(2\sqrt{2\ln 2})$. %$\sigma_f$ that can be varied in the range $\sigma_f \in [0.02{\rm , \ } 0.1]$ Hz. 
%$\Delta f$ is the full width of this spectrum at half maximum.

The control parameters are the initial carrier wave steepness $\epsilon_{\eta} \in[0.08{\rm , \ } 0.14]$ and the bandwidth $\Delta f \in [0.047{\rm , \ } 0.24]$ Hz that are varied in these ranges. A linear frame supports an array of 12 resistive wave probes at distance $x$ from the wavemaker with $x=3.4$, 5.2, 7, 8.9, 10.7, 12.6, 14.5, 16.3, 18.2, 20.1, 24.9, to $29.8$ m, respectively. Their vertical resolution is approximately 0.1 mm and their frequency resolution is close to 20 Hz, the sampling frequency being 250 Hz. A few additional probes are also present normal to the basin length to check that the wave field presents no significant evolution along the transverse direction. The surface elevation, $\eta(t)$, is recorded at each probe during $\mathcal{T}=2000$ s. We checked that the computed wave spectrum has converged statistically by computing it over the first and the second half of the signal duration $\mathcal{T}$. Note also that $\mathcal{T}$ is much greater than the autocorrelation time of the noise, $\mathcal{T} \gg (\Delta f) ^{-1}$. Typically, the wave amplitudes are of the order of few cm. Viscous dissipation is weak at these frequencies, the main damping mechanism being the beach, which absorbs more than 90\% of the incident energy. In a first approximation, our experimental setup can be thus considered satisfying the conservative hypothesis of  NLSE. Note that the NLSE hypothesis of slow time modulation is also verified experimentally ($0.05 \leq \Delta f/f_0 \leq 0.26$).

%used in the large wave basin in Ecole Centrale de Nantes (50 m long $\times$ 30 m wide $\times$ 5 m deep). The carrier wave frequency is fixed to $f_0=0.9$ Hz corresponding, using Eq.\ (\ref{displin}), to a wavelength of $\lambda_0=1.9$ m in a deep regime ($k_0 h \approx 16$). The period $T_0=1/f_0$ of the carrier wave in the basin is thus 1.1 s. Our statistics, at each measurement location, is averaged over the total experiment duration of 2000 s. This carrier wave is slowly perturbed with a noise of narrow frequency spectral bandwidth $\Delta f$ ( $\Delta f/f_0 \leq 0.1$). The initial carrier wave steepness $\epsilon_{\eta}\in[0.08{\rm , \ } 0.14]$ and $\Delta f \in [0.02{\rm , \ } 0.1]$ Hz are varied in these ranges.

\section{Time scale estimations}
Before describing the results, let us compute some typical time scales of this experiment. 
Consider first the parameters involved in the NLS Eq.\ (\ref{NLS}): The group velocity is $c_g=\omega_0/(2k_0)=0.87$ m/s, the dispersive parameter is $P=-0.06$ Hz m$^2$, and the nonlinear one is $Q=-30$ Hz/m$^2$. Second, we estimate the dispersive and nonlinear propagation times. Using Eq.\ (\ref{Tnl}), and $\epsilon=\sqrt{2}\epsilon_{\eta}$, the nonlinear propagation time reads
\begin{equation}
T_{nl}=1/(\omega_0\epsilon_{\eta}^2) \in [9 {\rm , \ } 27.6]{\rm \ s,}
\end{equation}
corresponding to a nonlinear length of $L_{nl} \in [7.8 {\rm , \ } 24]$ m. Using Eq.\ (\ref{Tlin}), $\Delta \omega/\omega_0=\Delta k/(2k_0)$, and $\Delta \omega=2\pi\Delta f$, the dispersive time scale reads
\begin{equation}
T_{lin}=\omega_0/(\Delta \omega)^2 \in [2.5 {\rm , \ } 64.6]{\rm \ s,}
\end{equation}
corresponding to a dispersive length of $L_{lin} \in [2.2 {\rm , \ } 56]$ m. Both $L_{nl}$ and $L_{lin}$ fit the length of the basin. Note that, for a fixed carrier wave frequency $\omega_0$, varying the initial wave steepness $\epsilon_{\eta}$ modifies $T_{nl}$, whereas varying the spectral bandwidth $\Delta \omega\equiv 2\pi \Delta f$ modifies $T_{lin}$. Using Eq.\ (\ref{Tratio}), the propagation time ratio is inferred as $T_{lin}/T_{nl}=\epsilon_{\eta}^2/(\Delta \omega/\omega_0)^2$. In the following, we define the square root of this quantity as the parameter quantifying the nonlinearity-to-dispersion ratio,  
\begin{equation}
\tau \equiv \sqrt{T_{lin}/T_{nl}}=\epsilon_{\eta}/ (\Delta \omega/\omega_0) {\rm \ \cdot}
\end{equation}
The nonlinearity ($\epsilon_{\eta}$) and dispersion ($\Delta \omega/\omega_0$) are controllable parameters in this experiment. To observe coherent structures such as solitons governed by NLSE, nonlinear and dispersive effects have to be balanced (i.e. $\tau \sim 1$). The parameter ranges are thus similar, as evidenced by the values of $\tau\in [0.3 {\rm , \ }  2.6]$.

%When the steepness increases, one finds $T_{lin}/T_{nl}\in {\color{red} [2.3 {\rm , \ } 7.1]}$ for {\color{red}$\Delta f=0.047$} Hz and $T_{lin}/T_{nl}\in {\color{red} [0.09 {\rm , \ } 0.28]}$ for {\color{red}$\Delta f=0.24$ Hz}. To observe coherent structures such as solitons, nonlinear and dispersive effects has to be balanced{\color{red}, i.e. $T_{lin}/T_{nl}\sim 1$}. This is reached for {\color{red}intermediate} values of our narrow frequency bandwidth. 
Since $\tau=BFI_{\omega}/\sqrt{2}$, the modulation instability, at the most unstable wavenumber, occurs for a pure monochromatic wave when $\tau > 1/2$. The instability occurs when $\Delta f< 2f_0\epsilon_{\eta} \in [0.14 {\rm , \ } 0.25]$ Hz. Note that the amplitude of the most unstable perturbation grows, during its propagation over a distance $L$, at most by a factor $\exp[(\omega_0\epsilon_{\eta})^2 2L/g]\simeq 3$ \cite{BenjaminFeir67,Melville82}. Finally, balancing directly the dispersive and nonlinear terms (i.e. third and fourth ones) in Eq.\ (\ref{NLS}) leads to the typical length $L_{sol}$ of envelope solitons \cite{Annexe4}
\begin{equation}
L_{sol}=\frac{\sqrt{2}\arcsech{(1/2)}}{k_0\epsilon_{\eta}} {\rm \ ,}
\label{Lcoh}
\end{equation}
as found in Eq.\ (\ref{Lsol}), $L_{sol}$ being the full width at half maximum of a $\sech$ pulse. 
This leads to solitons of typical length of 5 meters ($L_{sol} \in [4 {\rm , \ } 7.2]$ m) and duration of 5 seconds ($T_{sol}=L_{sol}/c_g \in [4.6 {\rm , \ } 8.3]$ s). Solitons have thus typically 4 to 8 carrier periods. The distance between the first and last probes in the basin is $L_{max}=26.5$ m. It is thus possible to follow such structures on propagation distances up to roughly 7 times its size, at best. To sum up, Table \ref{Table1} shows the different scales of the problem for parameter sets used in the experiments.

\begingroup
\squeezetable
\begin{table}[h]
\begin{tabular}{|c|c|c|c||c|c|c|c|}
\hline
$\epsilon_{\eta}$  &   $T_{lin}$ (s)   & $T_{nl}$ (s) & $\tau$ & $L_{lin}$ (m)   & $L_{nl}$ (m) & $L_{sol}$ (m) & $\frac{L_{max}}{L_{nl}}$\\
                            & $\frac{\omega_0}{(\Delta \omega)^2}$ & $\frac{1}{\omega_0\epsilon^2_{\eta}}$ & $\frac{\epsilon_{\eta}}{(\Delta \omega/\omega_0)}$ & $\frac{g}{2(\Delta \omega)^2}$ & $\frac{1}{2k_0\epsilon_{\eta}^2}$ & $\frac{1.86}{2k_0\epsilon_{\eta}}$ & \\
\hline
0.08 & $10.3$ & $27.6$ & $0.61$  & $9$ & $24$ & $7.14$ &$1.10$\\
\hline
0.12 & $10.3$ & $12.3$ & $0.92$ & $9$ & $10.6$ & $4.76$ & $2.48$\\
\hline
0.14 & $10.3$ & $9$ & $1.07$ & $9$ & $7.8$ & $4.08$ & $3.38$\\
\hline
\end{tabular}
\caption{Theoretical time and length scales for three different $\epsilon_{\eta}$ at fixed $\Delta f=0.05$ Hz. Carrier wave: $T_0=1.1$ s, $\lambda_0=1.9$~m.}
\label{Table1}
\end{table}
\endgroup

%{\color{blue} A terme, il faudrait peut-être :\\
%- augmenter $\Delta f\sim 0.15 - 0.18$ Hz pour avoir $T_{lin}/T_{nl} \sim 0.2$ (éviter l'instab. modul. pour $T_{lin}/T_{nl} > 0.5$) sans diminuer la cambrure sinon $L_{max}/L_{nl}<1$.\\
%- diminuer $\Delta f\sim 0.08 - 0.09$ Hz pour éviter d'avoir des $T_{lin}$ trop grand. Est ce que $f_0$ peut changer expérimentalement?\\
%- manips à $T_{lin}/T_{nl}=1$ en variant simultanément $\epsilon_{\eta}$ et $\Delta f/f_0$. Est ce que $f_0$ peut changer expérimentalement?\\
%- mettre des sondes plus loin??} 

  \begin{figure}
   \centering
  \includegraphics[width=19cm]{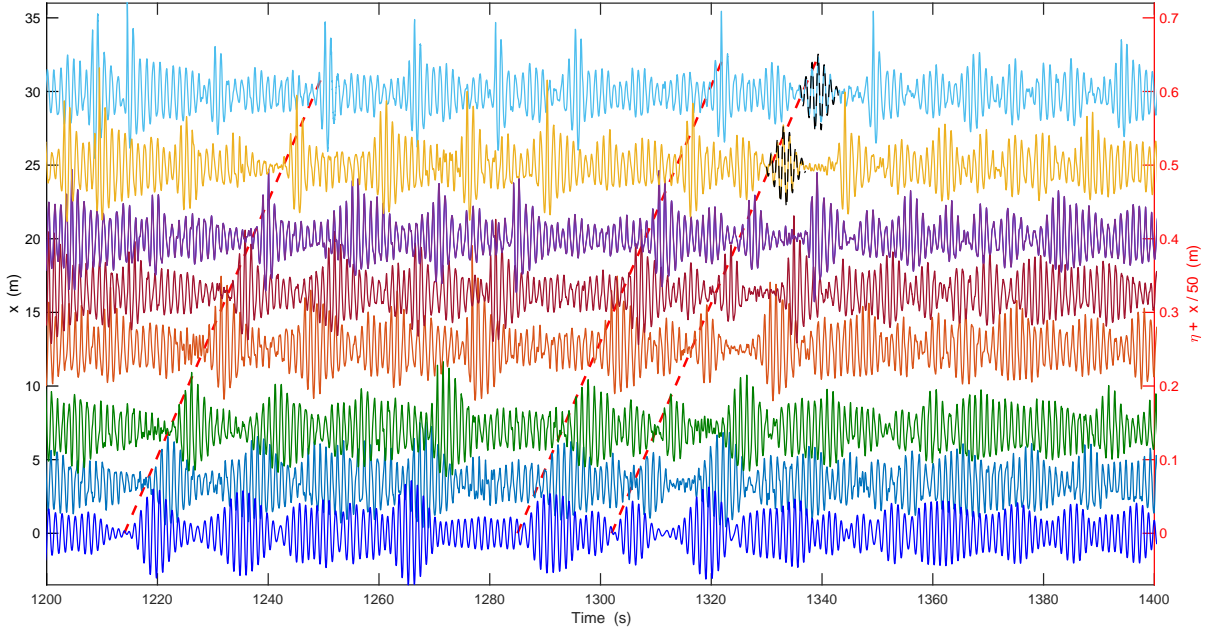}%{Fig2N.eps}
  \caption{Experimental evolution along the basin length, $x$, of the wave height $\eta(t)$ recorded at different probe locations (from bottom to top: $x=0$, 3.4, 7, 12.6, 16.3, 20.1, 24.9, and $29.8$ m). $\tau=0.44$ ($\epsilon_\eta = 0.08$ and $\Delta f = 0.16$ Hz).
  %$\sigma_f = 0.07$ Hz and $T_l/T_{nl}=0.19$. 
  Dashed (red) lines have a slope corresponding to the group velocity $c_g$. Dashed (black) line on top of the two upper curves: theoretical shape of the envelope soliton of Eq. (\ref{soliton}). }
  \label{etavstime}
  \end{figure}
  
% \begin{figure}
% \centering
% \includegraphics[width=6cm]{TlTnlPhaseRaid4.eps}
% \caption{Phase diagram of coherent structures in the ($T_{lin}$, $T_{nl}$) space. Crosses correspond to detected solitons. Solid line has a unity slope. $x = 29.8$m. Color code: violet (60), Bleu foncé (50), Vert (40), Bleu clair (30), Orange (20), rose(10). seuil écart-type vitesse verticale 4}
%  \label{PhaseSpace}
%  \end{figure}
  
%Test06_02 : cambrure=0.08, sigma_f=0.07Hz, Tnl=27.6s, Tl = 5.27 s, Tl/Tnl=0.19
%Test06_06 : cambrure=0.1, sigma_f=0.07Hz, Tnl=17.7s, Tl = 5.27 s, Tl/Tnl=0.3
%Test06_10 : cambrure=0.12, sigma_f=0.07Hz, Tnl=12.28s, Tl = 5.27 s, Tl/Tnl=0.43
% pour la figure Nombsolitons, j'utilise en plus Test06_01 : cambrure=0.08, sigma_f=0.1Hz, Tnl=27.6s, Tl = 2.58 s, Tl/Tnl=0.09

\section{Experimental results}
\begin{figure}
\centering
\includegraphics[height=5.5cm]{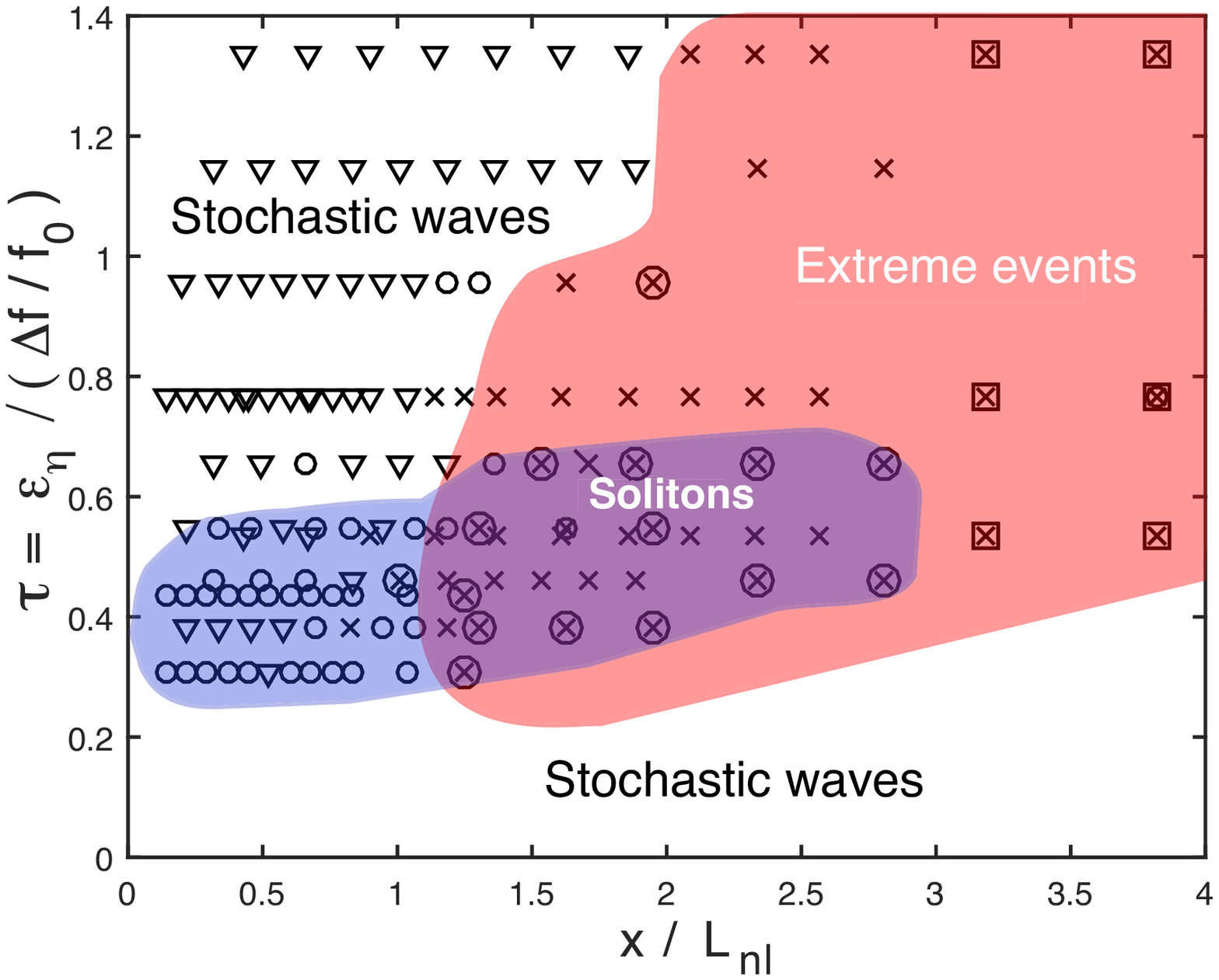}%{DiagPhase4-1N-1.eps}
  \includegraphics[height=5.8cm]{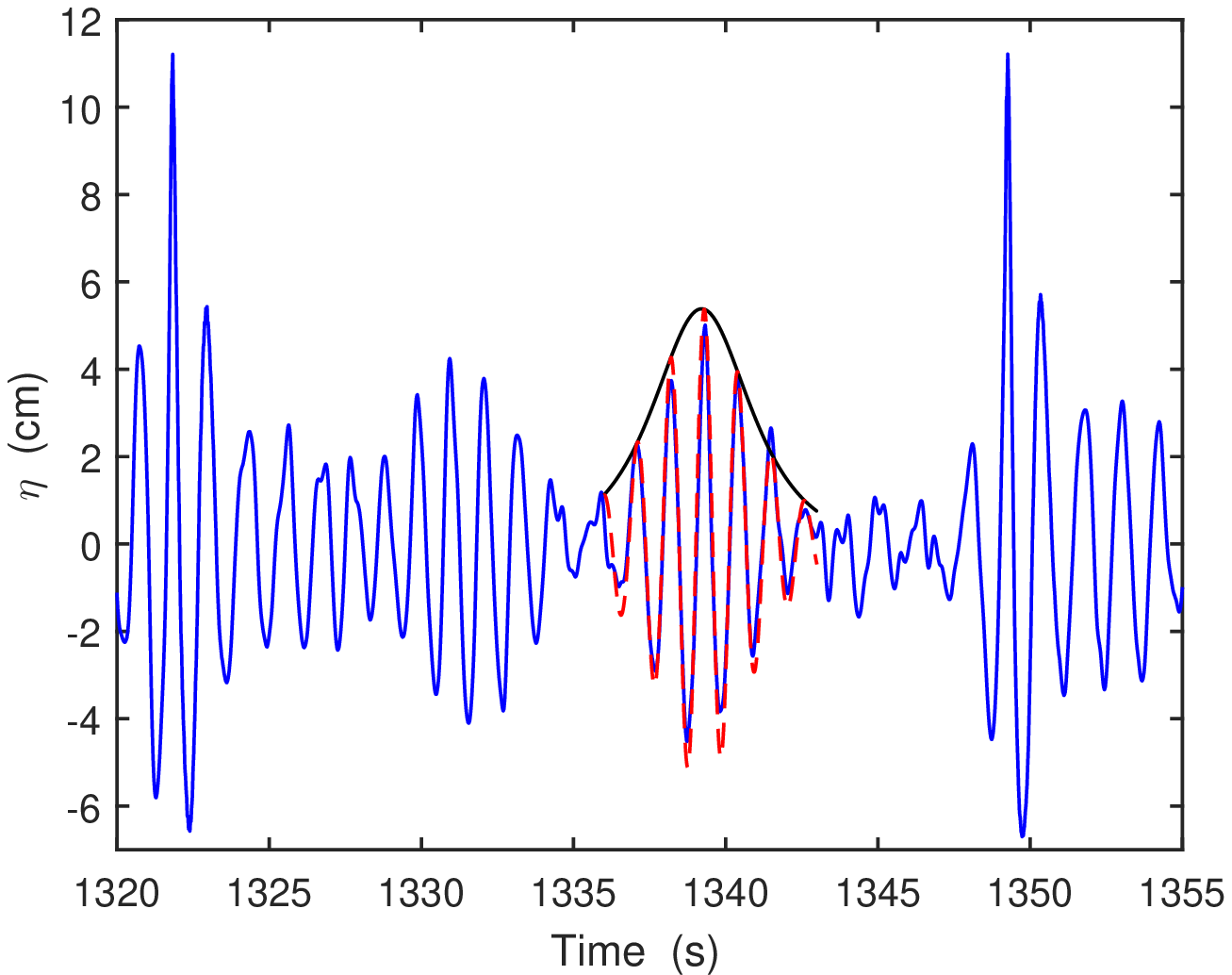}%{Fig3N-1.eps}
\caption{Phase diagram of stochastic waves (white area) coexisting with envelope solitons (violet area) or/and extreme events (orange area) as a function of the nonlinearity-to-dispersion ratio $\tau$ and the dimensionless distance $x/L_{nl}$.  Each performed run is displayed by a symbol corresponding to the observation of ($\triangledown$) only stochastic waves, ($\circ$) envelope solitons and ($\times$) extreme events coexisting with stochastic waves. ($\square$) indicates the coexistence with few spilling breakers (less than 10\%) in time series. Right: Typical envelope soliton detected at $x = 29.8$ m (zoom in of a part of the top curve in Fig.\ \ref{etavstime}), $\tau$=0.44 ($\epsilon_\eta = 0.08$, $\Delta f = 0.16$ Hz).
%$T_{lin}/T_{nl}$=0.19 ($\epsilon_\eta = 0.08$, $\sigma_f = 0.07$ Hz). 
($-$) Detected envelope. ($--$) Theoretical shape of NLS envelope soliton from Eq.~(\ref{soliton}). }
\label{ExSol}
\end{figure}

We generate unidirectional sinusoidal waves (of frequency $f_0$) subject to a slow and random phase modulation of the carrier ($\Delta f/f_0 < 0.26$). Figure\ \ref{etavstime} shows the temporal evolutions of the wave height, $\eta(t)$, recorded by the probes located at different distances $x$ from the wavemaker. Close to the wavemaker, the signal $\eta(t)$ is reminiscent from the forcing one standing for propagating wave packets of gentle amplitudes. As the distance increases, two main observations are reported. First, fronts of some wave packets steepen strongly leading to extreme events of large amplitude in the signal (see top curve in Fig.\ \ref{etavstime}). To quantify this strong steepening of the wave front, we arbitrary define an event to be extreme when its local slope $|d\eta/dt| > 4\sigma_{d\eta/dt}$, with $\sigma_{d\eta/dt} \equiv \sqrt{\langle(d\eta/dt)^2\rangle}$ the rms value of the wave slope (see~\S \ref{extreme}). During their propagation, other wave packets develop into solitons and then propagate with no deformation (see two top curves). These pulses are found to be well described by the envelope soliton profile of Eq.\ (\ref{soliton}) with no fitting parameter (once its maximum amplitude is fixed) - see superimposed dashed lines in Fig.~\ref{etavstime}. Finally, other wave packets in the signal spread gently during their propagation due to dispersion. All these wave packets propagate with the linear group velocity $c_g$ (see red dashed line), the nonlinear correction being less than 1\% for this chosen wave steepness in Fig.\ \ref{etavstime}. In Fig. \ref{ExSol}(left), we report all the experimental runs on a phase diagram showing the coexistence of stochastic waves with envelope solitons or/and extreme events as a function of the nonlinearity-to-dispersion ratio $\tau$ and the dimensionless distance $x/L_{nl}$. The emergence of extreme events occurs after roughly one nonlinear length scale of propagation. Envelope solitons arise only in an area where nonlinearity and dispersion are weak (but finite), and of the same order of magnitude as expected from NLSE. Indeed, when the steepness is too weak, or the modulation spectral width too large, no solitons are observed. Note that, according to the value of $\epsilon_{\eta}$, the last probe is located in this diagram at different values of $x/L_{nl}$ since $L_{nl}$ depends on $\epsilon_{\eta}$. Superimposed symbols, for the same set of parameters, mean that different coherent structures coexist in a same time series. Two nearby symbols obtained for two different sets of parameters, mean that different behaviors are detected in the two corresponding time series. For $\tau > 1.5$, extreme events are no longer observed since the spectral width is too small to significantly modulate the carrier wave [not shown in Fig. \ref{ExSol}(left)]. Note  also that after $3L_{nl}$ of propagation few spilling breakers occur in time series (less than 10\% of extreme events). Thus, when nonlinearity and dispersion are weak and of the same order, we observe a superposition of many interacting coherent structures such as solitons and extreme events within a sea of random wave packets. We characterize below in detail these coherent structures.

\begin{figure}
 \centering
\includegraphics[height=5.5cm]{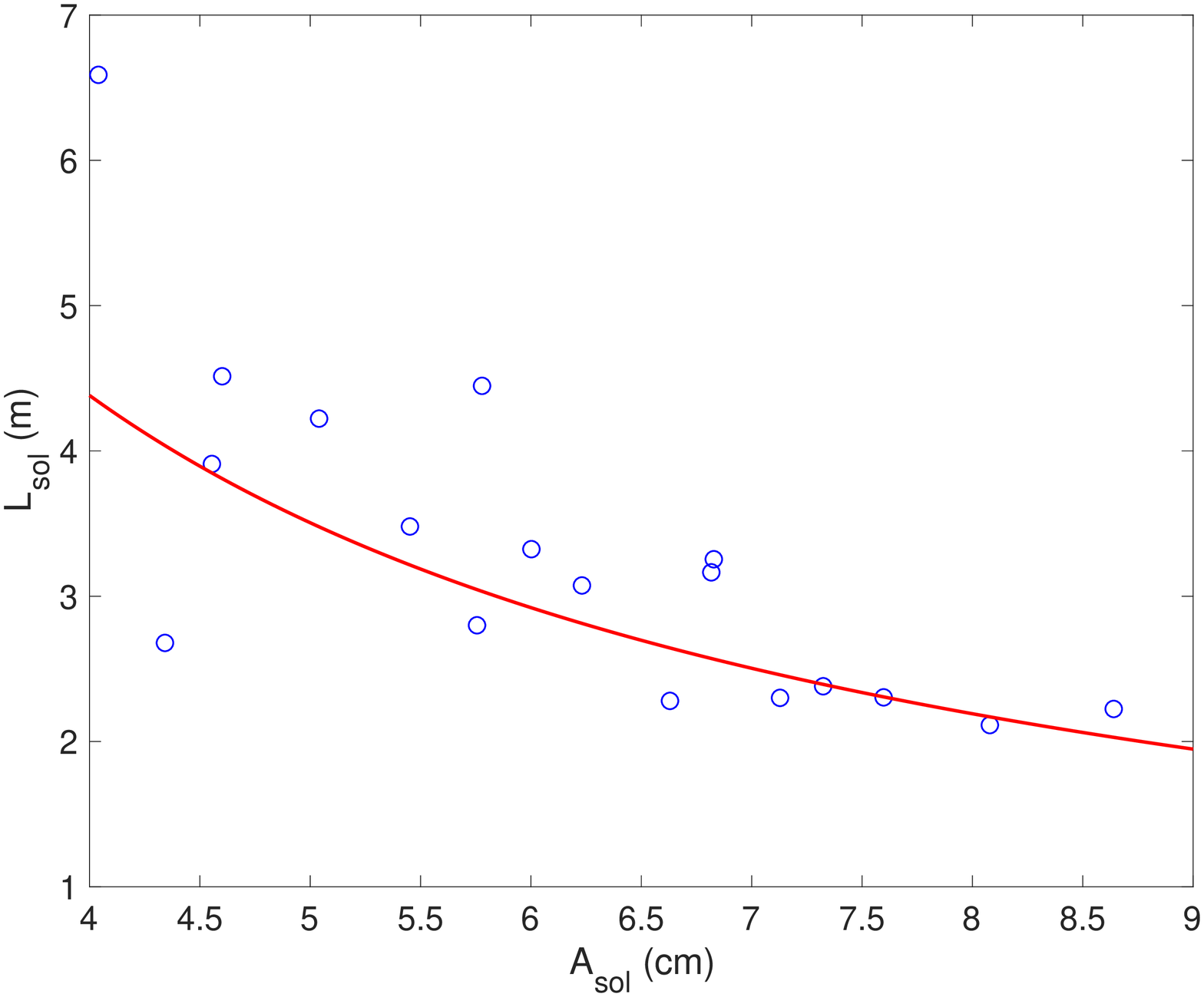}%{Fig4_leftN.eps}
\includegraphics[height=5.5cm]{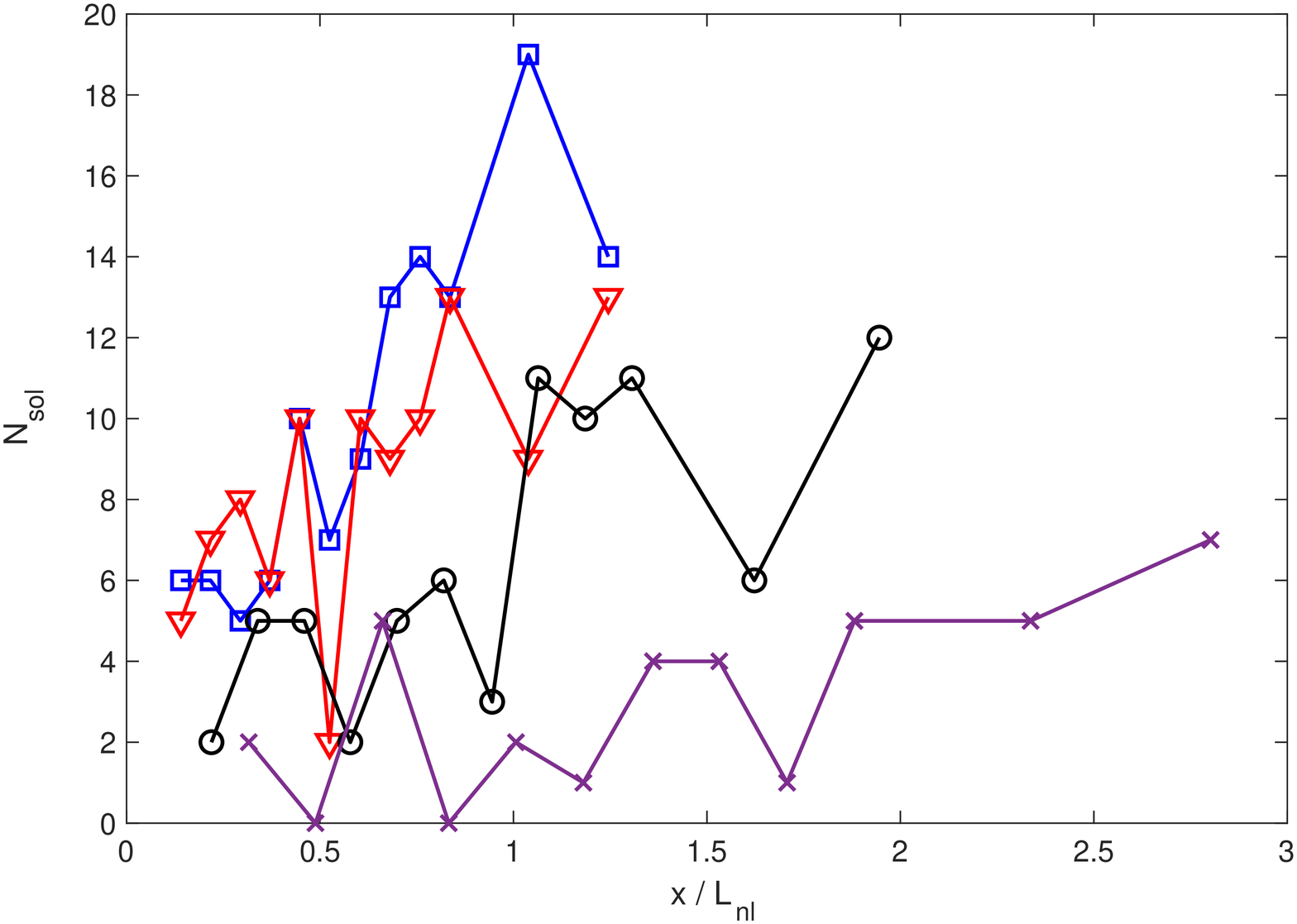}%{Fig4_rightN.eps}
\caption{Left: Full width (at half maximum) of detected solitons as a function of their maximum amplitude, $A_{\rm sol}$. $x=29.8$ m. $\tau$=0.55 ($\epsilon_\eta =0.1$, $\Delta f = 0.16$ Hz).  ($-$) Theoretical prediction from Eq.\ (\ref{Lsol}) with no fitting parameter. Right: Number of detected solitons as a function of the dimensionless distance $x/L_{nl}$ for different ratios $\tau=$ ($\square$) 0.30 ($\epsilon_\eta = 0.08$, $\Delta f = 0.24$ Hz), ($\triangledown$) 0.44 ($\epsilon_\eta = 0.08$, $\Delta f = 0.16$ Hz),  ($\circ$) 0.55 ($\epsilon_\eta = 0.1$, $\Delta f = 0.16$ Hz), and ($\times$) 0.65 ($\epsilon_\eta = 0.12$, $\Delta f = 0.16$ Hz).}
%$T_{lin}/T_{nl}=$ ($\square$) 0.09 ($\epsilon_\eta = 0.08$, $\sigma_f = 0.1$ Hz), ($\triangledown$) 0.19 ($\epsilon_\eta = 0.08$, $\sigma_f = 0.07$ Hz),  ($\circ$) 0.3 ($\epsilon_\eta = 0.1$, $\sigma_f = 0.07$ Hz), and ($\times$) 0.43 ($\epsilon_\eta = 0.12$, $\sigma_f = 0.07$ Hz). }  
\label{AspSol}
\end{figure}

%%%%%%%%%%%%%
 %%%SOLITONS 
\subsection{Solitons} 
A typical profile of a soliton within the signal $\eta(t)$ is shown in Fig.\ \ref{ExSol}(right) for the same experimental parameters as in Fig.\ \ref{etavstime}. Its profile is found in good agreement with the envelope soliton profile of Eq.\ (\ref{soliton}) with no fitting parameter (once its maximum amplitude is given). These solitons are observed with almost no deformation at least over two or three consecutive probes. To automatically detect the presence of envelope solitons within the temporal signal, we use a Hilbert transform and a thresholding method. The local maxima of the signal envelope are then detected and compared with the theoretical soliton profile, the fit being considered as successful when the correlation is better than 80\%. The widths $L_{sol}$ of the solitons detected within a single temporal signal are shown in Fig.\ \ref{AspSol}(left) as a function of their maximum amplitude $A_{sol}$. We found that the taller the soliton is, the narrower it is ; the data being well described by the NLSE prediction of Eq.\ (\ref{Lsol}) with no fitting parameter.
 
%  \begin{figure}
% \centering
%  \includegraphics[width=7cm]{/Users/efalcon/Recherche/NantesBassinHoule/CampagneC8Nov2016/TurbulenceIntegrableRes/figures_paper/Fig3.eps}%{C8_Test06_02ExSoliton1_191217}
%  %\includegraphics[width=7cm]{C8_Test06_02ExSoliton2}
%	\caption{Typical enveloppe soliton detected within the signal at $x = 29.8$ m. $T_{lin}/T_{nl}$=0.19 ($\epsilon_\eta = 0.08$, $\sigma_f = 0.07$ Hz). ($-$) Detected envelope. ($--$) Theoretical shape of NLS enveloppe soliton from Eq.~(\ref{soliton}) with no fitting parameter.}
%	\label{ExSol}
%\end{figure}

Figure \ref{AspSol}(right) shows the number $N_{sol}$ of detected envelope solitons as a function of the dimensionless distance $x/L_{nl}$ for different nonlinearity-to-dispersion ratios $\tau$. Regardless this value, $N_{sol}$ is found to increase with the distance showing thus that solitons are not present within the forcing, but emerge from the evolutions of wave packets during their propagation. Less solitons are detected as $\tau$ increases since the wave steepness $\epsilon_\eta$ has to be weak enough for NLSE to be valid. Note that 10 solitons are typically detected within a temporal signal, corresponding thus to a cumulated duration of 4\% of the latter. 

\begin{figure}
 \centering
 \includegraphics[height=6cm]{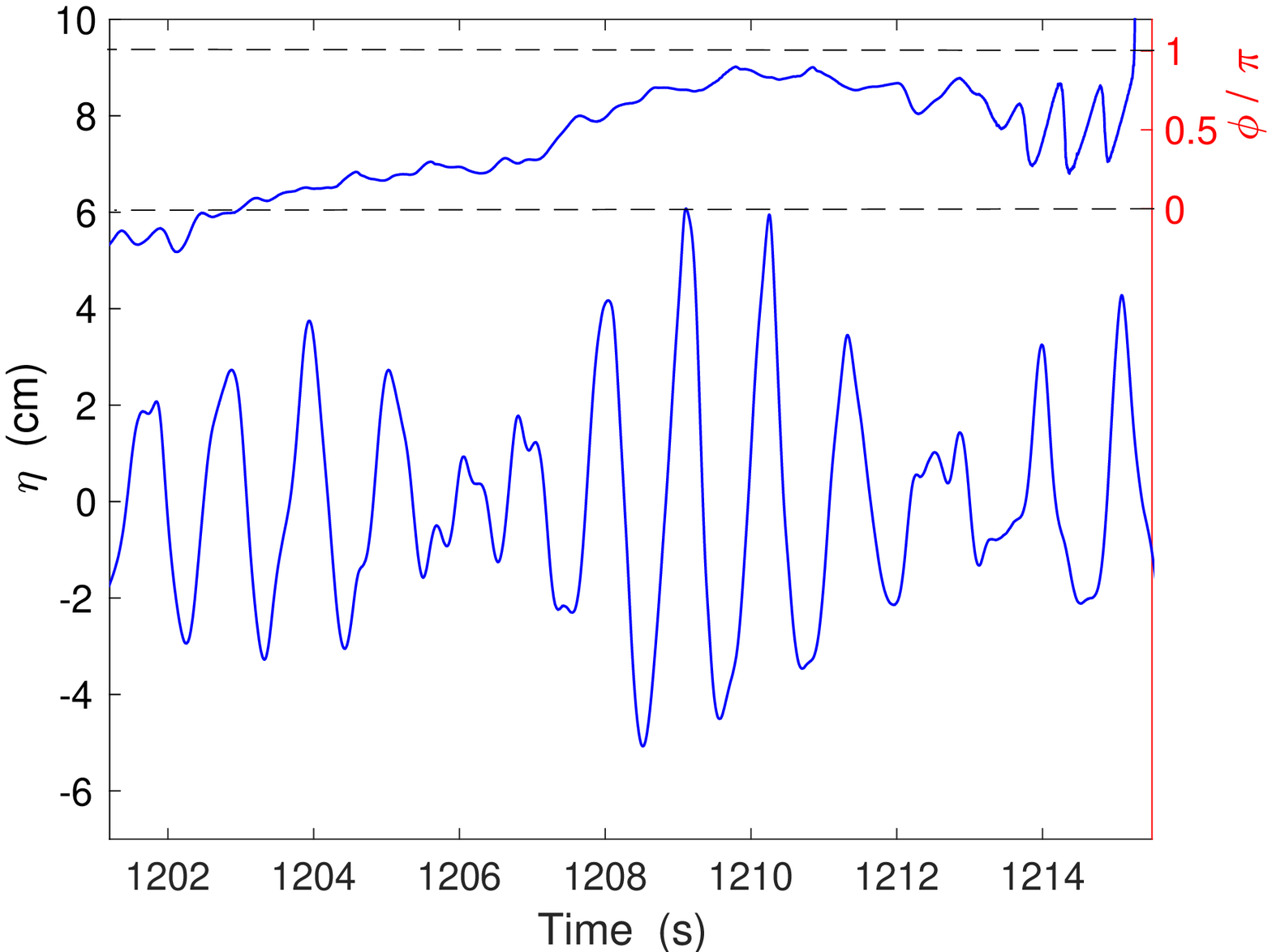}
\includegraphics[height=6cm]{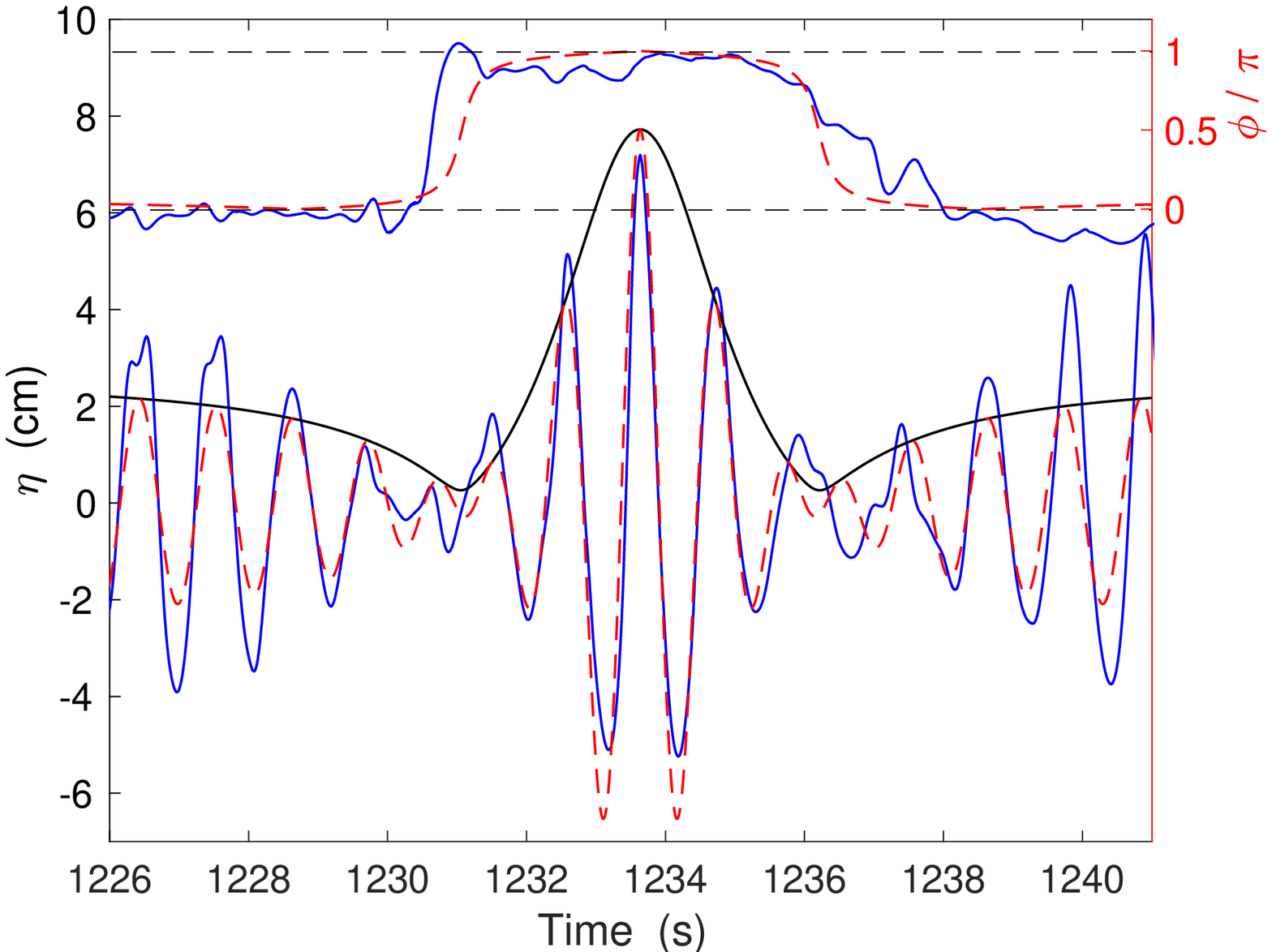}
\caption{Left: Wave height signal recorded close to the wavemaker at the first probe ($x= 3.4$ m). $\tau=0.3$ ($\epsilon_\eta = 0.08$, $\Delta f = 0.24$ Hz). Top: Rescaled phase $\phi/\pi$ of the signal. Right: Same part of the signal recorded at the second last probe ($x = 24.9$ m) showing a structure similar to a Peregrine soliton. Theoretical temporal profile ($--$) and envelope ($-$) of a Peregrine soliton from Eq.~(\ref{solitonP}) with $f_0=0.9$ Hz, $x=0$, and $k_0A_p=\epsilon_\eta$. Top: ($-$) Rescaled phase of the signal showing a $\pi$-jump at times where the envelope falls to zero as predicted ($--$) by Eq.~(\ref{solitonP}).}
\label{PeregrineFig}
\end{figure}

Another type of solitonic structure may appear in our time series. Indeed, Fig. \ref{PeregrineFig}(right) shows a pulse with a profile in good agreement with the Peregrine soliton of Eq.\ (\ref{solitonP}), both for its phase and envelope, with no fitting parameter (once its maximum amplitude is given). Although occurring much rarely than the envelope soliton in our time series, this structure similar to a Peregrine breather can be observed on a single probe emerging spontaneously from the noisy background. This structure, localized in time and space, is naturally not visible close to the wavemaker [see Fig.\ \ref{PeregrineFig}(left)]. Once it has been observed [see Fig. \ref{PeregrineFig}(right)], its amplitude recorded at the next probe decreases significantly. A signature of the Peregrine soliton is the $\pi$-jump of its phase across the zero amplitude domains separating the ``wings'' and the central lobe of the Peregrine soliton \cite{TikanPRL17}. We indeed report in Fig. \ref{PeregrineFig}(right) this characteristic $\pi$-phase jump at times where the amplitude falls to zero. As far as we know, this striking signature of the Peregrine soliton was reported only in optics \cite{TikanPRL17}, but not for water waves. In hydrodynamics, the Peregrine breather was observed when the wavemaker is forced by deterministic initial conditions (i.e. injecting the asymptotic Peregrine solution~\cite{Chabchoub11,ShemerPoF13,DongPRF18} that can be perturbated by an applied wind \cite{ChabchoubPoF2013}), by a periodic forcing (Stokes wave field) randomly noised \cite{ChabchoubProceed17}, or with a forcing consisting in the Peregrine solution embedded in a stochastic wave field \cite{ChabchoubPRL16}. To our knowledge, the emergence of a Peregrine soliton occurring from a fully stochastic forcing, as observed here, has been only reported in optics \cite{Walczak15,Suret16}. It is clear that further detailed investigations are needed to fully characterize this emerging localized structure, e.g. performing nonlinear spectral analysis \cite{RandouxPreprint18}, and to track it in a longer basin to reach a statistical quantification of its evolution and occurrences.

In the future, a local IST processing will be applied to our time series to precisely identify and classify the different types of coherent structures \cite{Randoux16b}. Note that other methods could be applied to find hierarchic solutions of NLSE such as the direct method (Hirota method), the B{\"a}cklund transformations or the Darboux transformations~\cite{InfeldBook,AkhmedievPRE09}. 

\begin{figure}
 \centering
\includegraphics[height=5.5cm]{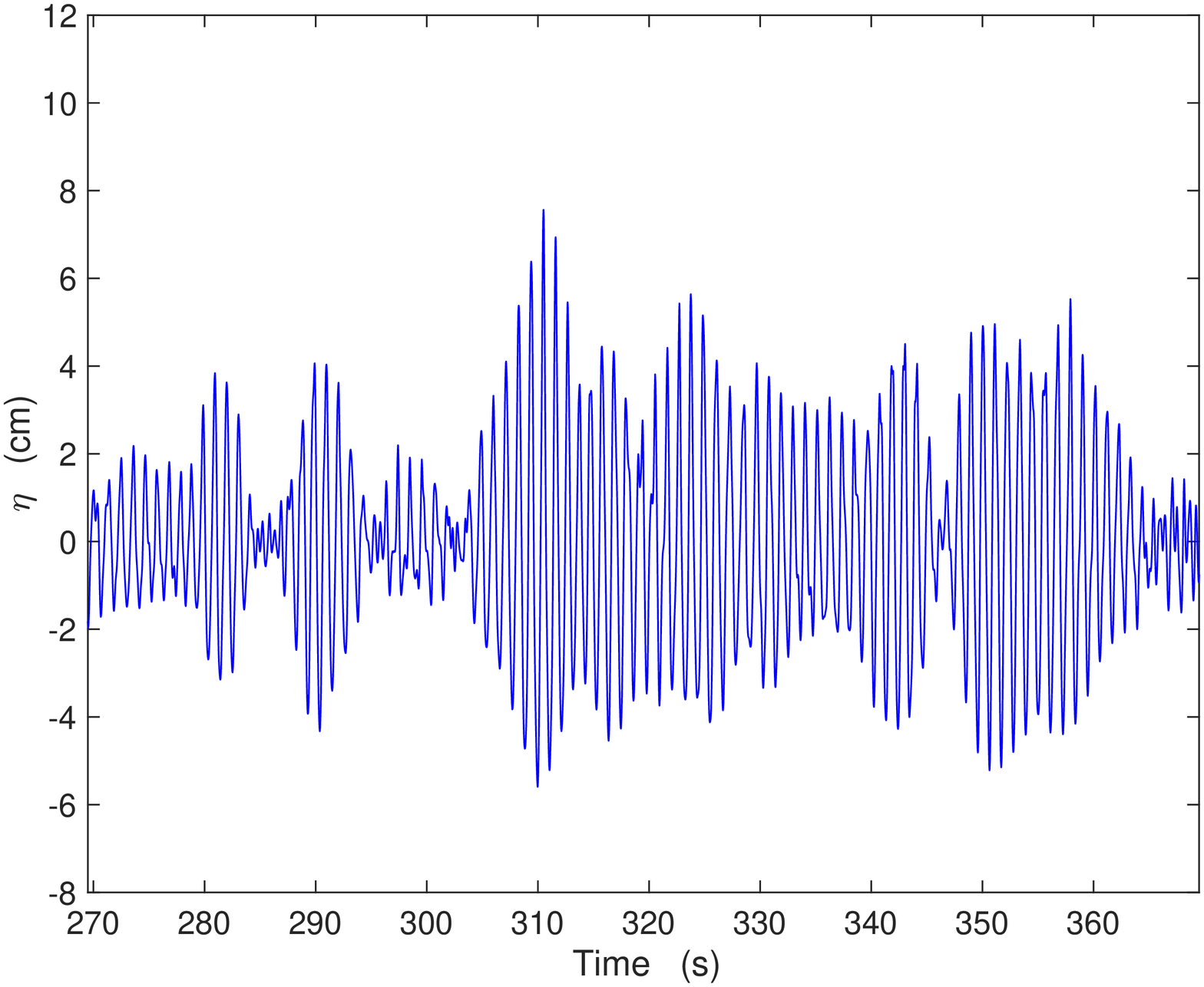}%{Fig5_leftN.eps}
\includegraphics[height=5.5cm]{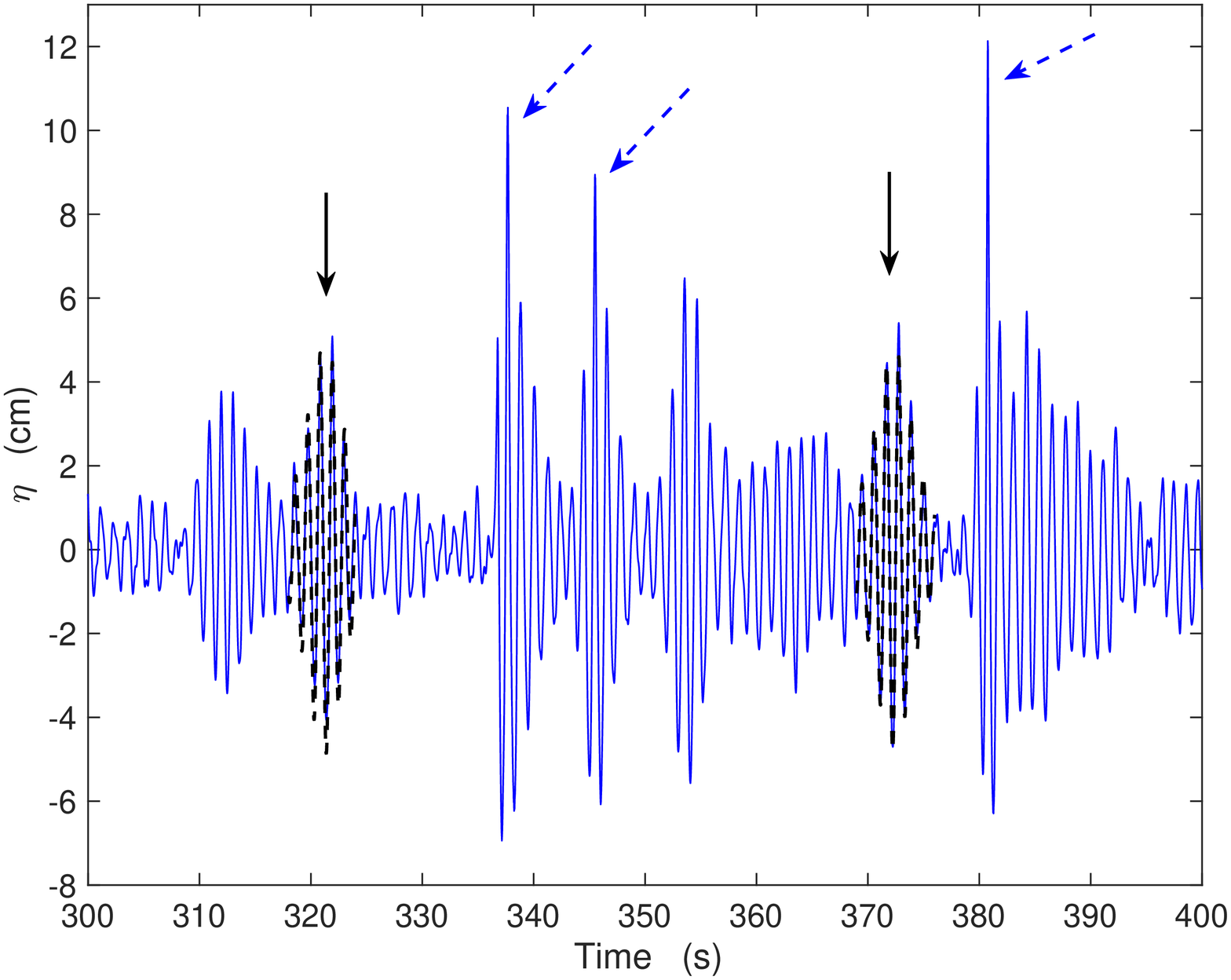}%{Fig5_rightN.eps}
\caption{Left: Wave height signal recorded close to the wavemaker at the first probe ($x= 3.4$ m). $\tau=0.55$ ($\epsilon_\eta = 0.1$, $\Delta f = 0.16$ Hz). Right: Same part of the signal recorded at the last probe ($x = 29.8$ m) showing coexistence of envelope solitons (see full arrows) and extreme events (see dashed arrows). ($--$) Theoretical shape of soliton from Eq.~(\ref{soliton}). }
\label{AspectSignal}
\end{figure}

Let us now have a look at the temporal evolution of the wave height, recorded at the first and the last probes, in the reference frame moving with the group velocity $c_g$. Close to the wavemaker [Fig.\ \ref{AspectSignal}(left)], the wave amplitude is slowly modulated leading to wave trains of erratic amplitudes and widths. Far from the wavemaker, three steep events of large amplitude have emerged (see dashed arrows) as well as two envelope solitons (see solid arrows) well described by the prediction of Eq.\ (\ref{soliton}). %Such a coexistence of solitons and extreme events within a sea of smaller incoherent waves is theoretically expected by the integrable soliton turbulence \cite{Zakharov09,Zakharov04}. 

%\begin{figure}
% \centering
%  \includegraphics[width=8.5cm]{NombSolitonsTlTnl121217.eps}
%  \caption{Number of detected solitons as a function of  dimensionless distance $x/L_{nl}$ for different ratios $T_{lin}/T_{nl}=0.09$ ($\square$), 0.19 ($\triangledown$), 0.3 ($\circ$), and 0.43 ($\times$). }
%  \label{NombSol}
%  \end{figure}

   %%%%%%%%%
 %Transition

  %%%%%%%%%
 %Evenement extremes
\subsection{Extreme events}
\label{extreme}
\begin{figure}
\centering
\includegraphics[height=5.7cm]{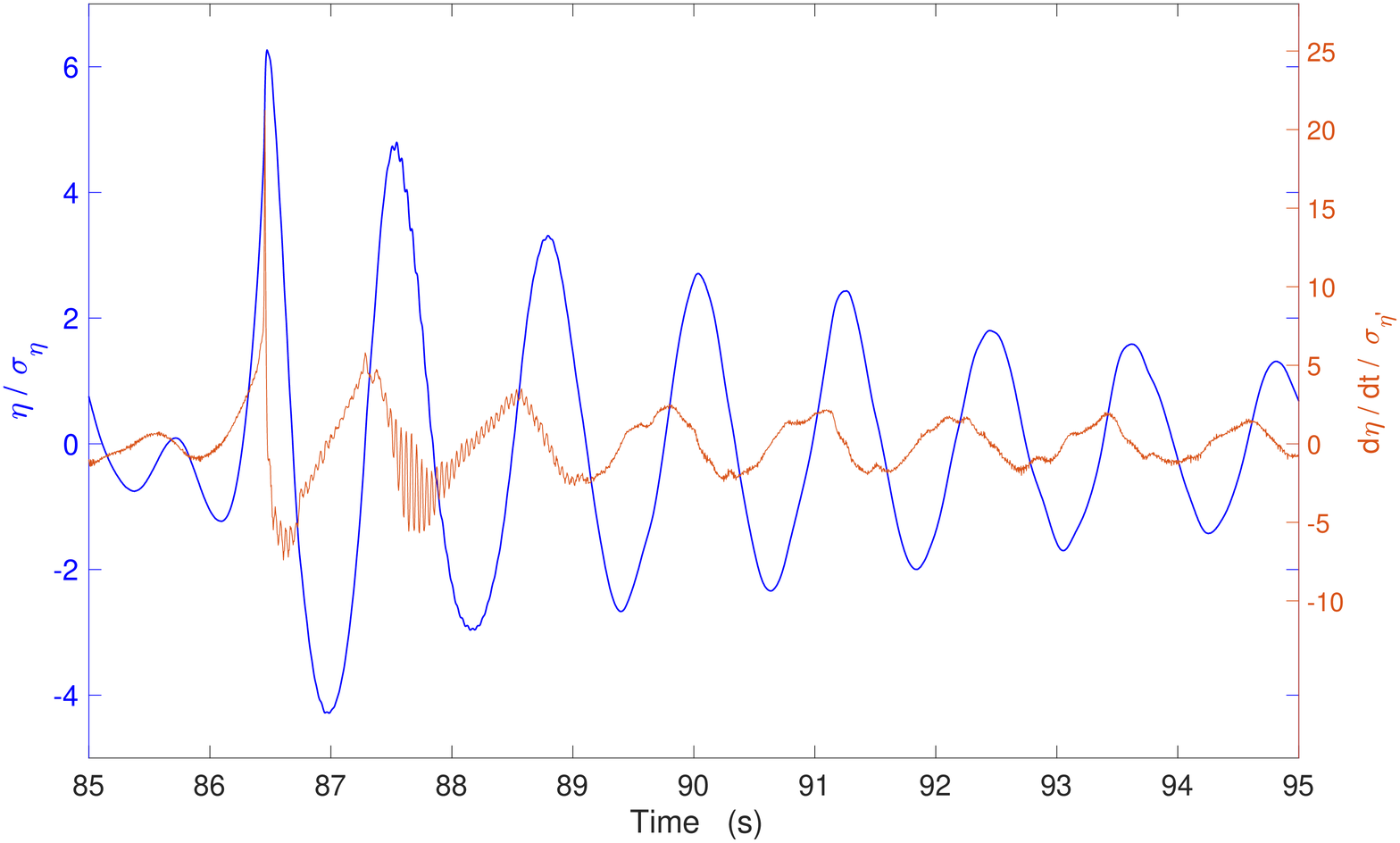}%{Fig6_leftN.eps}
\includegraphics[height=5.7cm]{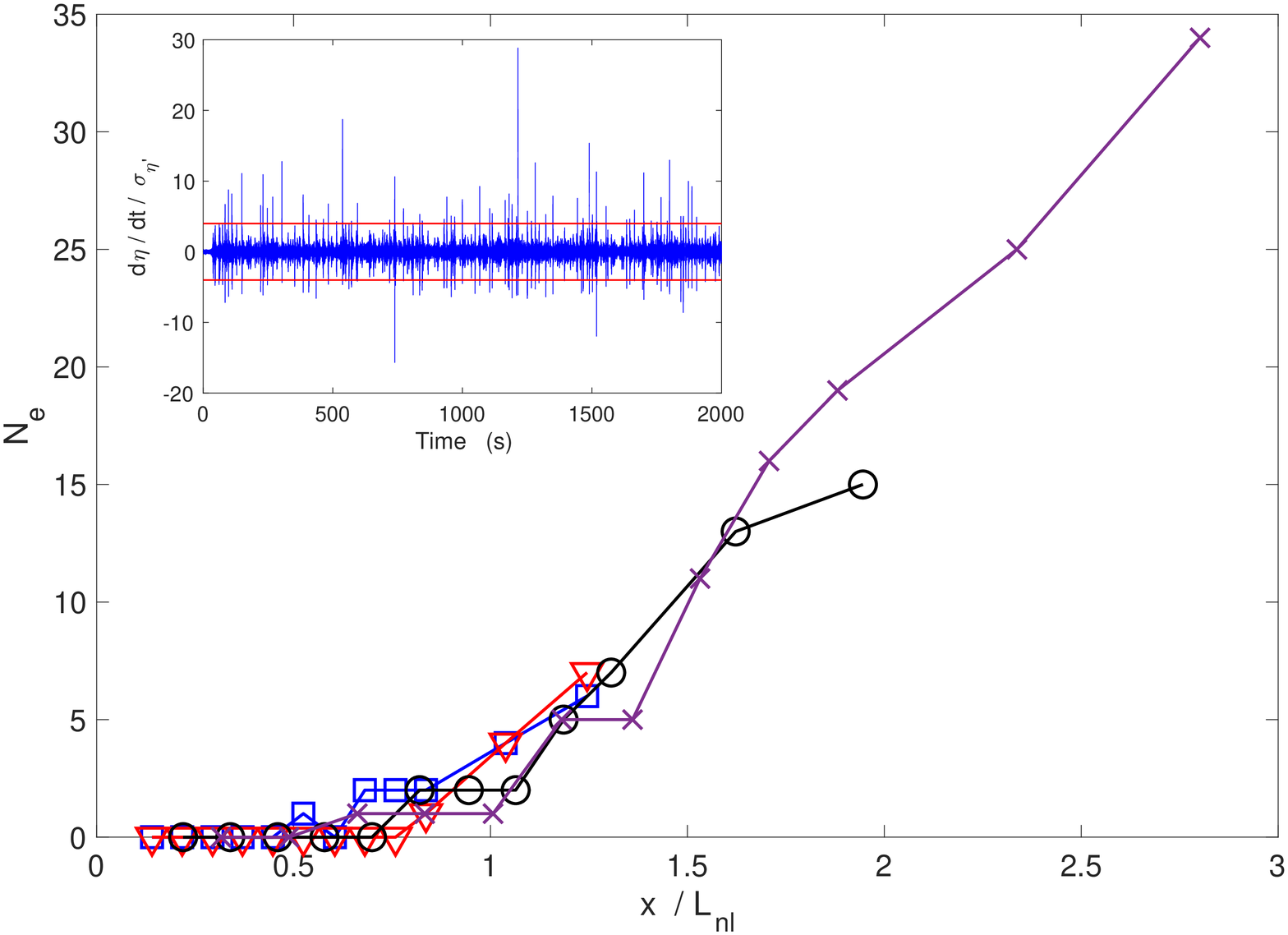}%{Fig6_rightN.eps}
\caption{Left: A typical extreme event as function of time at $x= 29.8$ m. Wavefront is the left-hand side. Normalized wave height, $\eta(t)/\sigma_{\eta}$ is on the left-hand side axis, and normalized wave local slope, $(d\eta/dt) /\sigma_{\dot{\eta}}$ is on the right-hand side axis. $\tau=0.44$ ($\epsilon_\eta = 0.08$, $\Delta f =0.16$ Hz). $\sigma_{\eta}=2.1$ cm. Right: Number $N_e$ of extreme events detected as a function of the dimensionless distance, $x/L_{nl}$, for different ratios $\tau$ (same symbols as in Fig.\ \ref{AspSol}). Inset: typical temporal signal of the wave local slope for $\tau=0.65$ ($\epsilon_\eta =0.12$, $\Delta f = 0.16$ Hz) at $x=29.8$ m. Solid lines correspond to $\pm 4\sigma_{d\eta/dt}$.}
\label{ExtremEvent}
\end{figure}

We characterize now the extreme events detected above. By zooming in on such a structure as in Fig.\ \ref{ExtremEvent}(left), one observes a very steep gravity wave front of very high amplitude [more than $6\sigma_{\eta}$ here]. This very steep propagative pulse followed by a slow decrease of the envelope, is thus highly asymmetrical with respect to time. This observation is magnified by superimposing on the same figure the wave local slope $d\eta/dt$ [computed from the differential of $\eta(t)$]. It shows an intense and short peak occurring on the forward face of the wave close to the maximum. After the main peak, a radiative tail follows over typically 5 to 10 periods. These steep events are found to occur randomly in the signal and have erratic amplitudes (see below).  The short oscillations of very small amplitudes visible on the wave slope signal is an experimental artifact due to the probe mechanical resonance ($\sim 20$ Hz) after the passage of the front. Finally, note that for high enough wave steepnesses ($\epsilon_\eta \geq 0.12$), less than 10\% of extreme events corresponds to the early stage of gentle spilling breakers (formation of a bulge in the profile on the forward face of the wave) \cite{Duncan99,Falcon10}. However, most of the results presented here are related to the dynamics of wave train steepening and solitons, for which dispersion and nonlinearity are of the same order of magnitude and weak enough to be described by NLSE.  

In order to quantify the number of extreme events, we arbitrary choose a criterion on the local wave slope, $|d\eta/dt|> 4\sigma_{d\eta/dt}$, instead of the usual one on the amplitude (4$\sigma_\eta$ or twice the significant wave height). Indeed, we want to characterize quantitatively the extreme events with very steep fronts, such as the one in Fig. 7(left), that contribute significantly to the high frequency part of the wave spectrum (see below). Note that 100\% of these detected events have an amplitude larger than $3\sigma_\eta$, and 70 to 85\% (depending on the forcing parameters) larger than 4$\sigma_\eta$.
%In order to quantify the number of extreme events, we arbitrary choose the criterion $|d\eta/dt| > 4\sigma_{d\eta/dt}$ to define an extreme event. 
As shown in the inset of Fig.\ \ref{ExtremEvent}(right), peaks of very high amplitudes occur randomly in the wave slope signal, most of them being larger than  $\pm 4\sigma_{d\eta/dt}$. Typically, the cumulated duration of these extreme events is 10\% of the signal duration. The number $N_e$ of extreme events detected with this thresholding method is shown in Fig.\ \ref{ExtremEvent}(right) as a function of the distance for different nonlinearity-to-dispersion ratio, $\tau$. $N_e$ is found to increase from zero with the distance showing thus that these extreme events results form the steepening and merging of the wave trains during their propagation. $N_e$ is also found to be independent on $\tau$ within our range, when rescaling the propagating distance, $x$, by the nonlinear length scale, $L_{nl}$, based on NLSE. The onset of occurrence of such steep coherent structures seems thus to be well described by NLSE. Moreover, $N_e$ increases linearly with this rescaled distance once the wave field propagated more than roughly one nonlinear propagation length scale. Finally, we checked that same qualitative results are found when varying the above thresholding criterion in the range ($\pm 3\sigma_{d\eta/dt}$,$\pm 6\sigma_{d\eta/dt}$).

%\begin{figure}
%\centering
%\includegraphics[width=8.5cm]{NraidTlTnl.eps}
%\caption{Number of extreme events detected as a function of  dimensionless distance $x/L_{nl}$ for different ratios $T_{lin}/T_{nl}$ (same symbols as in Fig.\ \ref{NombSol}). Inset: typical temporal signal of the wave local slope for $T_{lin}/T_{nl}=0.43$. $x=29.8$ m. $\epsilon_\eta =0.13$. $\sigma_f = 0.07$ Hz. Solid lines correspond to $\pm 4\sigma_{\dot{\eta}}$.}
%\label{NbExtremEvent}
%\end{figure}

  %%%%%%%%
  %%%Spectrogram
   \begin{figure}
   \centering
  \includegraphics[height=6cm]{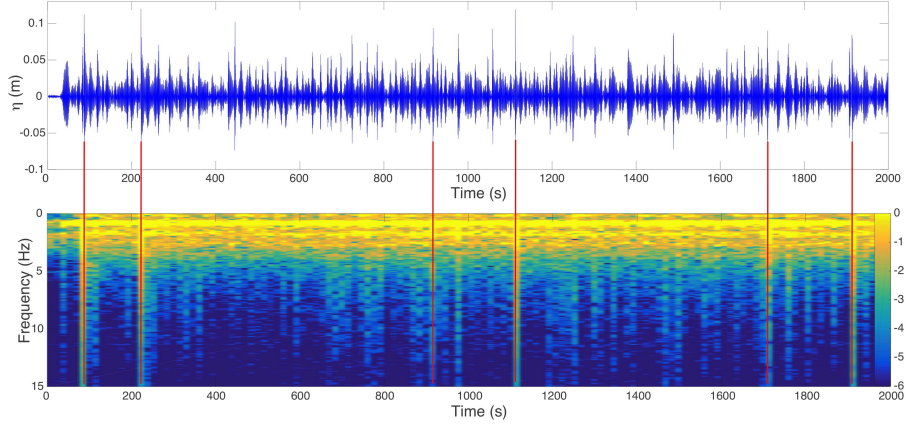}%{Fig7-1N.eps}
  \caption{Top: Temporal wave height signal for $\tau$=0.44 ($\epsilon_\eta = 0.08$, $\Delta f =0.16$ Hz). $x=29.8$ m. Bottom: Time-frequency spectrum $S_{\eta}(f,t)$ of the corresponding wave height $\eta(t)$. Color bar is a logarithmic scale of the spectrum amplitude. Large amplitude events of the wave signal correspond to maxima of spectrum at high frequencies (see solid lines).}
  \label{Tempsfreq}
  \end{figure}
    
  %%%%%%%%
  %%%Spectre
 \subsection{Wave spectrum}
 When dispersive and nonlinear effects are of the same order of magnitude, the above experimental results show the presence of solitons (solutions of the NLSE), emerging from the initial random forcing conditions, as well as strong steepening of some wave train fronts, both coherent structures occurring randomly in the incoherent wave field. To quantify the spectral content of such an erratic signal $\eta(t)$ as displayed in the top inset of Fig.\ \ref{Tempsfreq}, we compute its time-frequency spectrum $S_{\eta}(f,t)$. To wit, a short-time Fourier transform of $\eta(t)$ is computed by fast-Fourier transforms of overlapping windowed signal segments (using the Spectrogram function from Matlab software). The wave spectrum is thus reached at each time over a short time interval. The wave spectrum as a function of time and frequency is shown in Fig.\ \ref{Tempsfreq}. As expected, its main contributions are related to the random forcing band near $f_0$ and its corresponding harmonics ($nf_0$ with $n=2$ and 3). More interesting is the spectral signature of extreme events. Each intense peak within the wave signal gives a continuous high frequency contribution to the spectrum. Some of these similarities are emphasized by solid lines in Fig.\ \ref{Tempsfreq}. Extreme events thus contain high frequencies due to their steep profile.
 
   \begin{figure}
\centering
\includegraphics[height=6cm]{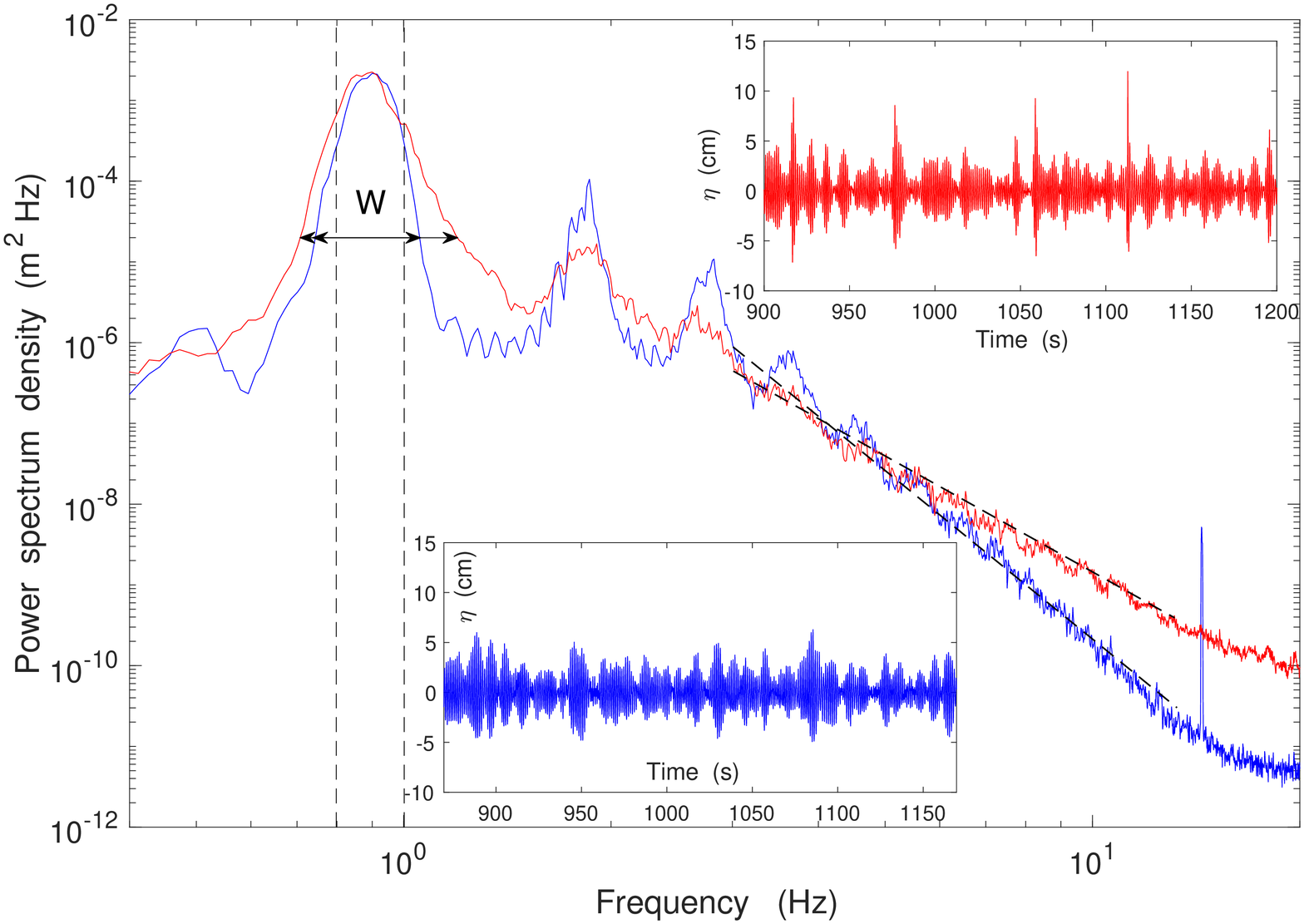}%{Fig8_leftNN.eps}
\includegraphics[height=6cm]{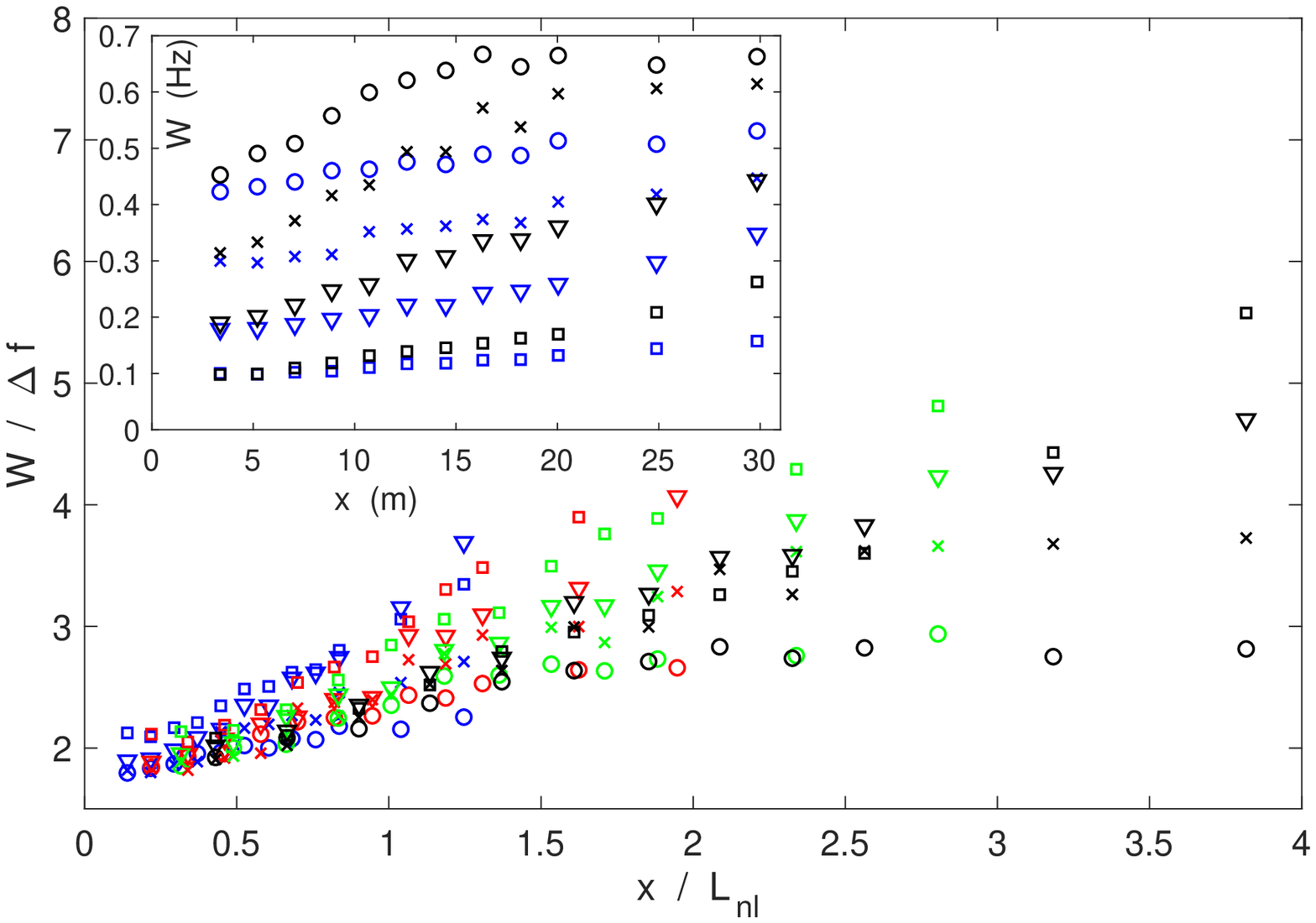}%{Fig8rightNNN.eps}
\caption{Left: Power spectrum density of $\eta(t)$ recorded close [$x=3.4$ m (blue)] and far [$x=29.8$ m (red)] from the wavemaker. Dashed lines have slopes $\alpha=-6.4$ (blue) and $\alpha=-4.5$ (red). $\tau=0.44$. Vertical dashed lines correspond to theoretical satellites of the Benjamin-Feir instability, $\pm c_gK_c/(2\pi)\simeq \pm 0.1$ Hz.  Insets show the corresponding temporal wave height signals close (bottom) and far (top) from the wavemaker. Right: Dimensionless width, $W/\Delta f$, of the main peak of the spectrum (at one hundredth of its maximum amplitude) as a function of $x/L_{nl}$. $\Delta f=$  ($\circ$) 0.1, ($\times$) 0.07, ($\triangledown$) 0.04, and ($\square$) 0.02 Hz. $\epsilon_\eta =$ (blue) 0.08, (red) 0.1, (green) 0.12, (black) 0.14. Inset: Unrescaled curve, $W$ vs. $x$. Same symbols as in the main figure.}
\label{spectrum}
\end{figure}
 
Figure \ref{spectrum}(left) shows the spectra averaged over time of a wave field recorded at the first probe, close to the wavemaker, and also at the last probe far from the wavemaker [see insets of Fig.\ \ref{spectrum}(left)]. Here again, close to the wavemaker, a discrete spectrum with main contributions related to the forcing domain near $f_0$ and its corresponding harmonics ($nf_0$ visible up to $n=5$). Far from the wavemaker, the high frequency components of the spectrum, as well as frequency domains between successive harmonics, have strongly increased to the detriment of harmonics amplitudes. It thus leads to a monotonic spectrum that is found to decrease as a frequency power law of the form $f^{-4.5}$. Extreme events emerging during the propagation [see top inset of Fig.\ \ref{spectrum}(left)] thus populate the high frequencies of the spectrum. Consequently, far enough from the wavemaker ($x/L_{nl} > 1.5$), nonlinear effects are sufficient to generate extreme events (resulting from the front steepening) that significantly contribute to the building of the high frequency part of the spectrum. Indeed, steep extreme events are known to be rich in harmonics. In the low-frequency part, the spectrum develops a visible asymmetry and a broadening of the main peak near $f_0$  with the propagation distance, as also observed in Ref. \cite{Shemer09}. Indeed, the width $W$ of the main peak increases linearly with the distance $x$ as shown in the inset of Fig. \ref{spectrum}(right). These data roughly collapse on a single curve by plotting the dimensionless width $W/\Delta f$ vs. the dimensionless distance $x/L_{nl}$ [see Fig. \ref{spectrum}(right)]. This broadening is known to be well described by NLSE contrary to the asymmetry that is captured by a higher (fourth) order extension of the NLSE (Dysthe model) to account for finite spectrum width \cite{ShemerPoF10}. 
 
\begin{figure}[h]
\centering
\includegraphics[height=6cm]{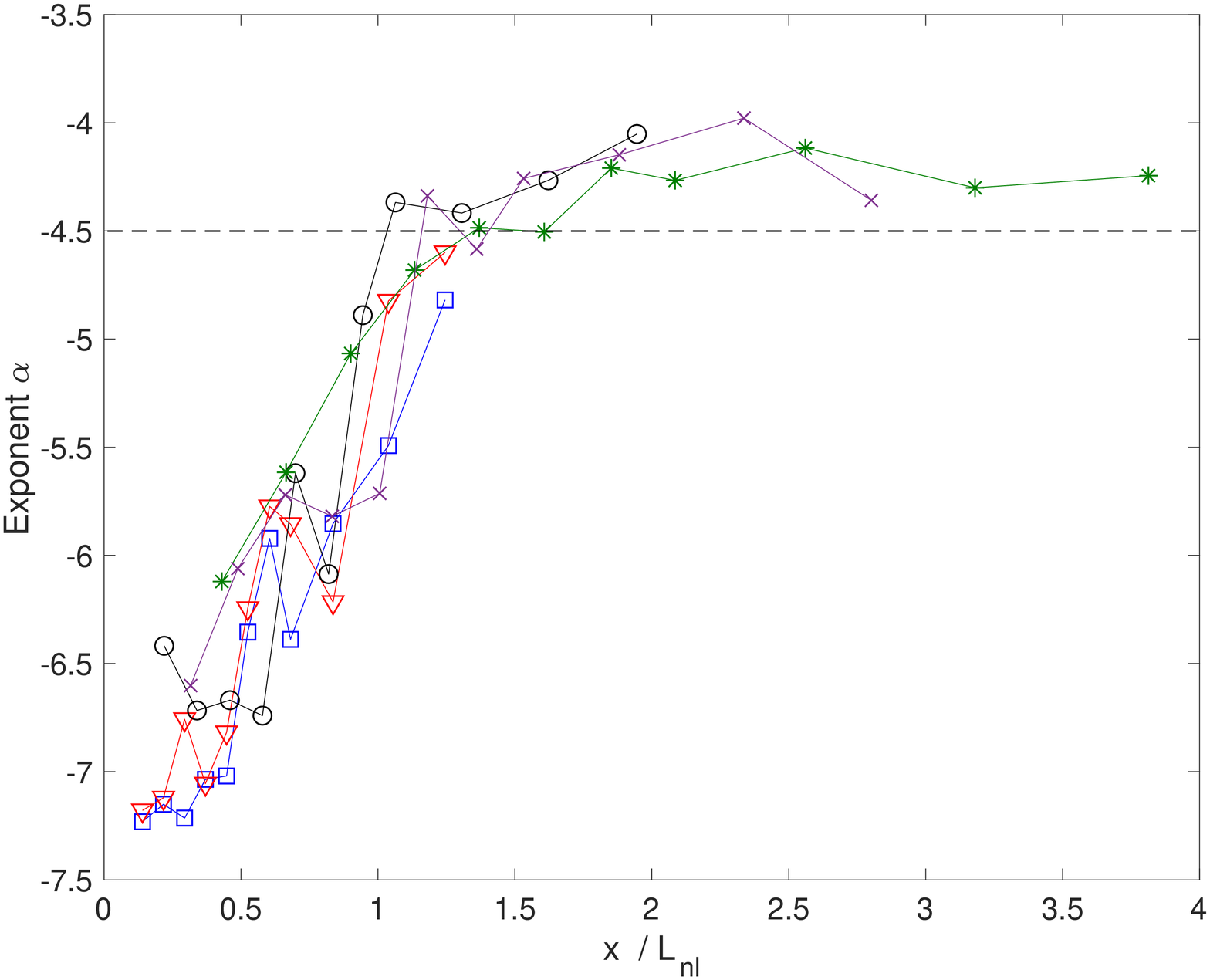}%{Fig9_left-1N.eps}
\includegraphics[height=6cm]{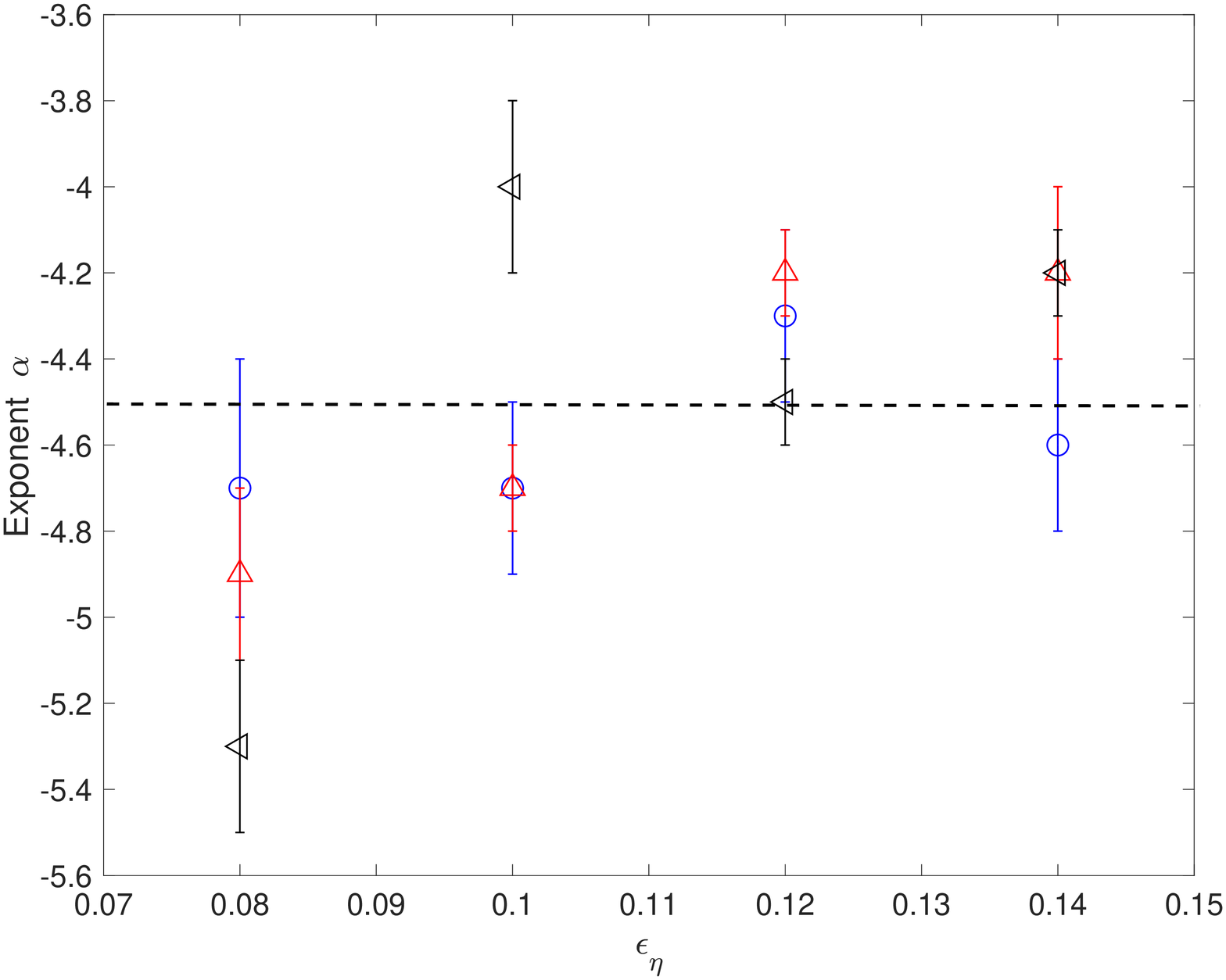}%{Fig9_rightNN.eps}
\caption{Left: Exponent $\alpha$ of the wave spectrum in $f^{\alpha}$ as a function of dimensionless distance $x/L_{nl}$ for different ratios $\tau$  (same symbols as in Fig.\ \ref{AspSol}). Right: Exponent $\alpha$ as a function of the wave steepness for different modulation bandwidth $\Delta f=0.09$ ($\circ$), 0.16 ($\bigtriangleup$) 0.24 ($\triangleleft$) Hz at $x/L_{nl}>1.2$. Dashed lines show the 1D wave turbulent prediction $-9/2$. }
\label{alpha}
\end{figure}
  
Let us now further characterize the high-frequency part of the spectrum. A frequency-power law spectrum $\sim f^{\alpha}$ is observed regardless the values of our parameters ($\epsilon_{\eta}$ and $\Delta f$) except for a too slow modulation ($\Delta f \leq 0.05$). The evolution of the spectrum exponent $\alpha$ with the distance for different $\tau$ ratios is shown in Fig.\ \ref{alpha}(left). $\alpha$ is found to be independent of $\tau$ in this range of parameters, when rescaling the propagating distance, $x$, by the nonlinear length scale, $L_{nl}$, based on NLSE. At short distances, random steepening of wave trains and solitons have not enough time to emerge in the wave field and the high frequency part of the spectrum is very steep in order to connect the noise level. When the wave field propagates over more than one and a half nonlinear propagation length scale ($x/L_{nl} > 1.5$), the exponent is roughly found to be constant near $\alpha \simeq -4.2$ as a result of the wave steepening as underlined above. Note that this power-law spectrum could be also ascribed as a signature of 1D gravity wave turbulence phenomenon. Indeed, the prediction of 1D unidirectional gravity wave turbulence is $\alpha=-9/2$ \cite{Dyachenko95,Zakharov04,Connaughton03}. %This prediction is compatible for the experimental spectra of wave fields recorded far enough from the wavemaker (typically when $x/L_{nl} > 1$). 
However, the use of a beach as an efficient damping mechanism inhibits the occurrence of resonant interactions driven by reflected waves. Moreover, the carrier wave propagates during roughly 40 periods until it reach the beach which is too short to develop nonlinear interactions required by wave turbulence. Besides, the high-frequency part of the experimental spectrum has been shown above to be a consequence of the strong steepening of wave trains that are not taken into account neither by weak turbulence,  nor by NLSE. Indeed, the numerical spectrum of 1D random wave field described by NLSE is exponential near the carrier frequency and display peaks near its harmonics \cite{ShemerPoF10}. %and the nonlinear coherent structures of NLSE are symmetric \cite{SuretRandouxPC}. 
Here, the 1D random wave field have a continuous power-law spectrum. The detected extreme events resulting from strong wave steepening carry intrinsically numerous harmonics. The observed power-law scaling thus arises probably from the slowly random frequency modulation of the harmonics (bound waves) of the wave field, that is known to generate continuous power-law spectrum between $f^{-5}$ and $f^{-4}$ \cite{Michel2018}.
%Moreover, these events are rich in harmonics (bound waves), and a slowly random frequency modulation of such 1D anharmonic wave trains is known to be able to generate continuous frequency power-law spectrum with exponent between $-5$ and $-4$ \cite{Michel2018}. 
In a limit case, if the extreme events tend to display very sharp wave-crests (cusps), and are assumed to propagate without deformation (i.e. $\omega \sim k$), the spectrum of such singularities is predicted to scale as $f^{-4}$ \cite{Kuznetsov2004}, not far from the experimental results. Finally, the exponent $\alpha$ is shown in Fig.\ \ref{alpha}(right) as a function of the forcing strength (the initial wave steepness $\epsilon_{\eta}$). $\alpha$ is found to be roughly constant, $-5 < \alpha < 4$, with respect to $\epsilon_{\eta}$ within the experimental estimation accuracy, showing thus its independence from $\epsilon_{\eta}$ for our range. To sum up, this power-law spectrum not described by NLSE arises probably from the random modulation of the harmonics (bound waves) of the carrier wave.

 \subsection{Wave field statistics}
The statistical properties of wave fields in integrable turbulence governed by focusing NLSE have been experimentally studied recently in optics, and show the emergence of heavy-tailed statistics \cite{Walczak15,Randoux16,Suret16}. In hydrodynamics, non Gaussian wave statistics have been observed experimentally during the propagation of unidirectional gravity waves forced with random initial conditions in a deep water regime \cite{Onorato04,Shemer09,ShemerPoF10,ShemerJGR10,ElKoussaifiPRE18} as predicted theoretically by using NLSE with random initial forcing \cite{JanssenJPO03}. However, it has not been related to the integrable turbulence in the hydrodynamics case. Here, we discuss the wave field statistics obtained in our experiment.

\begin{figure}
\centering
\includegraphics[height=5.5cm]{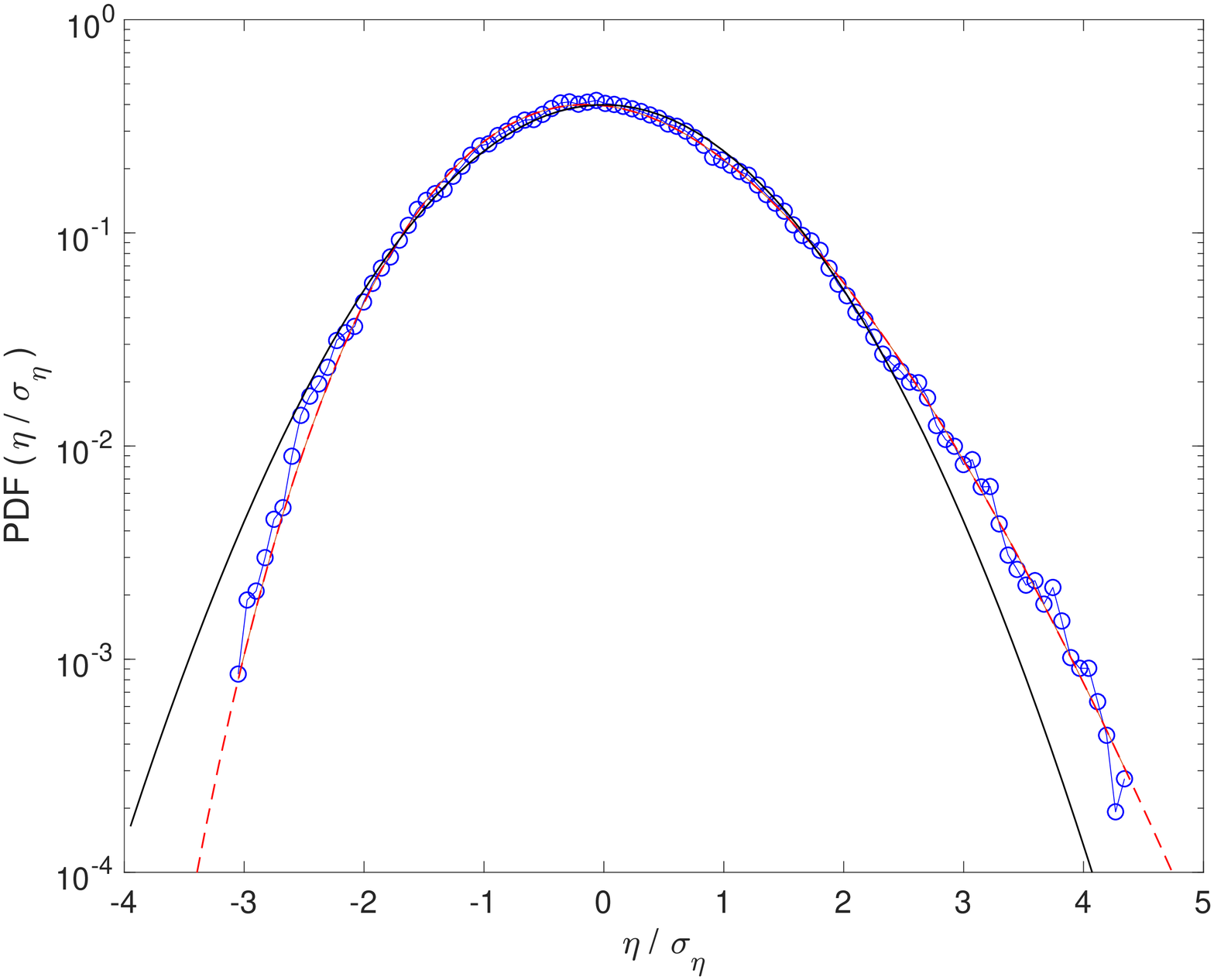}%{Fig10_leftN.eps}
\includegraphics[height=5.5cm]{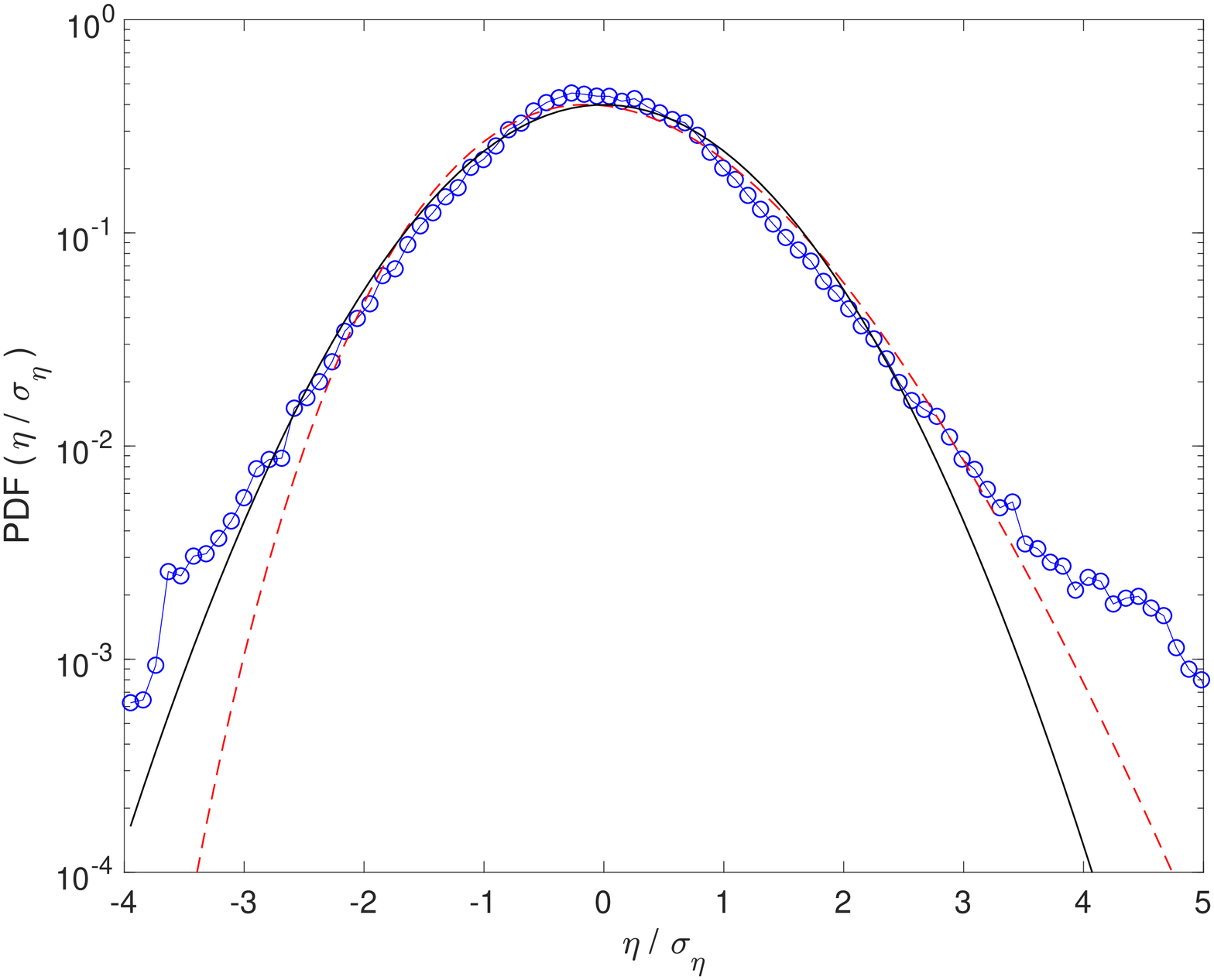}%{Fig10_rightN.eps}
\caption{Probability density function (PDF) of normalized wave height, $\eta/\sigma_\eta$, recorded at the first (left) and last (right) probe ($x=3.4$ m and $x=29.8$ m, respectively). $\tau$=0.55 ($\epsilon_\eta=0.1$, $\Delta f =0.16$ Hz). Solid lines display a Gaussian of zero mean and unit standard deviation. Dashed lines show a Tayfun distribution for a wave steepness of 0.1.}
\label{PDFTayfun}
\end{figure}

Figure\ \ref{PDFTayfun} shows the typical probability density function (PDF) of normalized wave height, $\eta/\sigma_{\eta}$, recorded at the first and last probes. Close to the wavemaker, the PDF is found to be asymmetric since large crests are more probable than deep troughs as a consequence of the nonlinear effects that are well described by Tayfun distribution (first nonlinear correction to a Gaussian) \cite{Tayfun80} (see dashed lines). This PDF asymmetry is routinely observed in laboratory experiments \cite{Onorato04,Falcon07} and in oceanography \cite{Forristall00}. Far from the wavemaker, the PDF departs from the Tayfun distribution near $3\sigma_{\eta}$ meaning that high amplitude events are more probable. Such heavy-tailed distribution has also been already reported experimentally \cite{Onorato04,Shemer09,ShemerPoF10,ShemerJGR10,Hassaini17} and could be related to rogue wave formation in ocean \cite{SotoCrespo16,Onorato01,Slunyaev06,Onorato04,Osborne05,Islas05}.

%Cette observation a déjà été faite par Onorato et al.  en 2004 qui interprètent ces événements de grande amplitude comme le résultat de l'instabilité de Benjamin-Feir. L'instabilité engendrerait des ``modes de breather'', des ondes non-linéaires qui croissent atteignent un maximum, puis reviennent à leur état initial de manière récurrente. Toutefois, pour prouver l'existence de tels modes, il faudrait pouvoir mesurer le champ d'ondes de manière spatiotemporelle, des mesures locales ne suffisent pas.

\begin{figure}
\includegraphics[height=5.5cm]{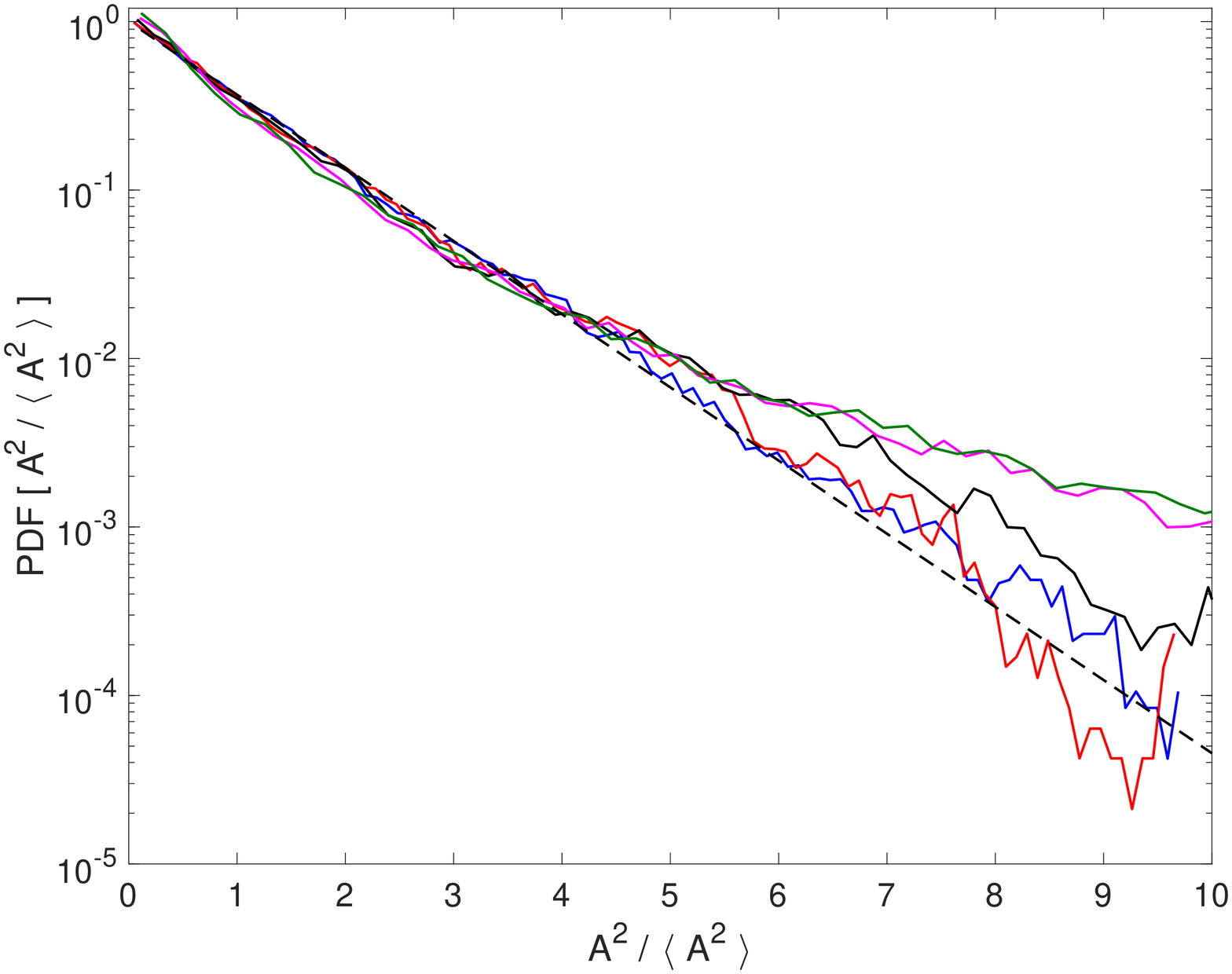}%Fig11_leftN.eps
\includegraphics[height=5.5cm]{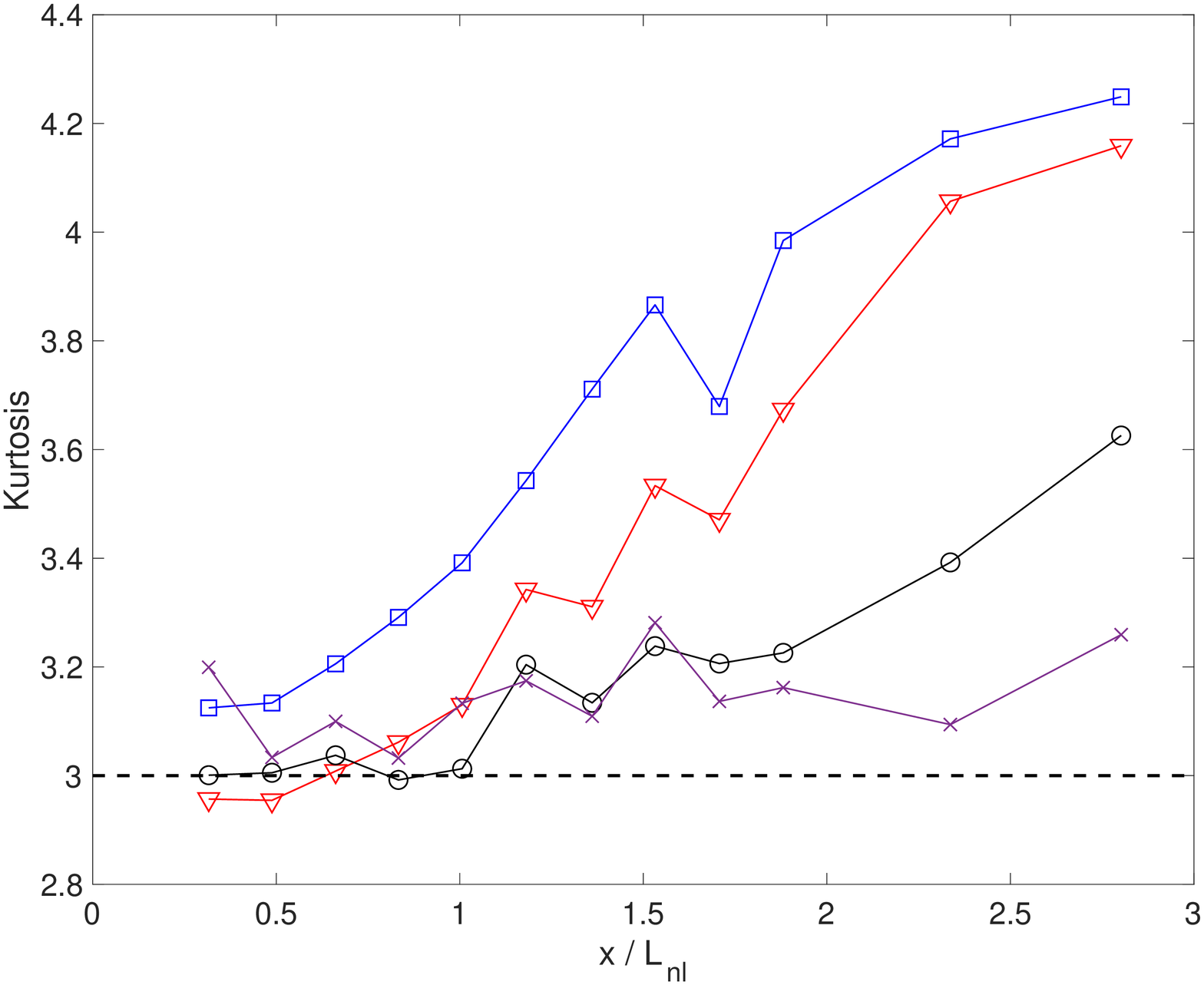}%Fig11_rightN.eps
  \caption{Left: Probability density function (PDF) of the square of the wave envelope, $A^2/\langle A^2\rangle$, at different distances $x/L_{nl} = 0.22$, 0.58, 0.82, 1.62 and 1.94 (from bottom to top). $L_{nl}=15.34$ m. $\tau=0.55$ ($\epsilon_{\eta}=0.1$, $\Delta f =0.16$ Hz). Dashed line: exponential (Rayleigh) distribution. Right: Kurtosis of $\eta(t)$ as a function of the dimensionless distance, $x / L_{nl}$, for constant $\epsilon_{\eta}=0.12$ and different $\Delta f=0.05$ ($\times$), 0.09 ($\circ$), 0.16 ($\triangledown$), and 0.24 Hz ($\square$) (i.e $\tau=2.29$, 1.15, 0.65, and 0.46 respectively). Dashed line displays Gaussian value ($K=3$).} %Inset: $K-3$ versus $(\Delta f/f_0) / \epsilon_{\eta}$ for $x / L_{nl}=2.94$. Solid line has a unity slope.}
  \label{PDFSondes}
\end{figure}

In optics, the statistics of the power fluctuations of light are measured (i.e. the square of the wave envelope) instead of the wave displacement. To be able to compare with these results, we compute the Hilbert transform of $\eta(t)$ to obtain the envelope $A(t)$. Note that if the statistics of random independent fluctuations, saying $\eta(t)$, follows a normal law, then an exponential (Rayleigh) distribution results for the wave envelope $A(t)=|\eta(t)|$ or for the ``power'' $A^2$ \cite{Evans00}. The PDF of the square of the wave envelope, $A^2/\langle A^2 \rangle$ is plotted in Fig. \ref{PDFSondes}(left) at different propagation distances. $\langle \cdot \rangle$ stands for a temporal average. As the wave field moves away from the wavemaker, the PDFs evolve from an exponential distribution (plotted in dashed-line) to a heavy-tailed distribution. For instance, power fluctuations 10 times greater than the mean power have a probability, far from the wavemaker, 30 times greater than the one close to the wavemaker. More precisely, we find that for $x/L_{nl} > 1.6$, the wave system reaches a statistical stationary state in which the PDF no longer changes with distance [see top two curves in Fig. \ref{PDFSondes}(left)], the power spectrum of waves being also independent of the distance [see Fig. \ref{alpha}(left)]. Similar results for the distance independent PDF have been observed in experiments with optics fibers governed by the focusing NLSE of Eq.\ (\ref{NLS}) as well as in numerical simulations of this equation in the context of integrable turbulence \cite{Walczak15,Randoux16,Suret16}. This statistical stationary state is stated to be determined by the interaction of coherent nonlinear structures \cite{Randoux16}. However, the mechanisms in integrable turbulence that lead to the establishment of this stationary state with such statistical properties independent of distance (or time) are an open question.

We compute now the skewness, $S\equiv \langle \eta^3\rangle/\langle \eta^2\rangle^{3/2}$, and the kurtosis, $K\equiv \langle \eta^4 \rangle/\langle \eta^2 \rangle^{2}$, of the wave height statistics quantifying its asymmetry and its flatness, respectively. For a Gaussian distribution, one has $S=0$ and $K=3$. At small distance, $S$ is non-zero confirming the asymmetry observed on the PDFs. This asymmetry $S\simeq 0.3$ is found to be roughly constant regardless of the propagation distance, $x$ and the nonlinearity-to-dispersion ratio, $\tau$. Figure \ref{PDFSondes}(right) shows the Kurtosis as a function of the dimensionless distance, $x/L_{nl}$, for different modulation bandwidths $\Delta f$ (i.e. different $\tau$) at fixed initial steepness $\epsilon_\eta$.  Consistently with the PDF observations, $K$ is found to increase with the distance regardless the forcing parameters [either increasing $\Delta f$ or $\epsilon_\eta$ (not shown here) keeping the other one constant]. Similar observations have been done in \cite{Onorato04}. $K$ increases strongly once one nonlinear propagation distance is reached. This is consistent with the fact that, during the wave propagation, more and more coherent structures (such as strong steepening of the wave trains) are generated [see Fig. \ref{ExtremEvent}(right)] and interact with the residual random wave field. However, when the nonlinearity-to-dispersion ratio $\tau$ is of the order of 0.5-0.6 (a value for which solitons and extreme events coexist [see Fig.\ \ref{ExSol}(left)]), a beginning of saturation of $K$ is observed with distance for $x/L_{nl} > 2$ [see top curves in Fig.\ \ref{PDFSondes}(right)]. This regime in which statistical properties of waves become independent on the distance is consistent with the above PDFs observations, and is also in agreement with experiments performed in a much longer basin  showing that NLSE reproduces well this Kurtosis behavior \cite{Shemer09,ShemerPoF10,ElKoussaifiPRE18}. Most efficient initial conditions of the random wave field to form a sea state with numerous and intense extreme events (i.e. large $K$) is thus for a weak enough (but finite) dispersion (i.e. dimensionless spectral width) of the order of twice the nonlinearity (steepness) of the wave field. 

\section{Conclusion}
In nonlinear physics, when nonlinearity is comparable to or exceeds dispersion, different structures may appear, such as conservative (like solitons) or dissipative structures resulting of finite-time singularity of the nondissipative equations (like shocks, wavebreaking) \cite{Falkovitch06}. Identifying such structures and the role they play in determining different stationary statistical states remains to be investigated in most turbulent systems. %For such high level of nonlinearity, no analytic theory is yet available for ``strong'' wave turbulence \cite{Falkovitch06}. Let us now sum up different existing theories in order to compare them with our experimental results. 

Here, we report the experimental observation of a new statistical state for unidirectional propagation of gravity waves in a deep water regime where coherent structures coexist with smaller stochastic waves. Such a state is predicted theoretically by NLSE integrable turbulence \cite{Zakharov09}, but had never been observed so far in this context. The nonlinearity ($\epsilon_{\eta}$) and dispersion ($\Delta \omega/\omega_0$) are controllable in our experiment, and are chosen similar to be able to observe solitons governed by NLSE. The nonlinearity-to-dispersion ratio, $\tau \equiv \epsilon_{\eta}/ (\Delta \omega/\omega_0)$ is varied from 0.3 to 2.6. We have characterized the emergence, the property and the evolution of these nonlinear coherent structures (solitons and extreme events) within the incoherent wave background. The emergence of extreme events resulting from the strong steepening of wave train fronts occurs after roughly one nonlinear length scale of propagation (estimated from NLSE). Envelope solitons and Peregrine solitons are also observed emerging from the stochastic background. Solitons arise when nonlinearity and dispersion are weak (but finite), and of the same order of magnitude, as expected from NLSE. The numbers of envelope solitons and extreme events are found to increase all along the propagation. When the nonlinear distance of propagation is reached, the wave spectrum is found to scale at high frequencies as $\omega^{-4.5\pm 0.5}$. This scaling is robust regardless of the variation of our parameter ranges. Although, this spectrum scaling could be compatible with the prediction of 1D gravity wave turbulence in $\omega^{-9/2}$ \cite{Dyachenko95,Zakharov04,Connaughton03}, the transfer mechanism towards small scales is not due to wave interactions, but is shown in the spectrogram to be ascribed to the strong wave steepening leading to the presence of extreme events. Since the latter carry numerous harmonics, this power-law scaling arises probably from the slowly random frequency modulation of the harmonics (bound waves) of the wave field that is known to generate continuous power-law spectrum between $\omega^{-5}$ and $\omega^{-4}$ \cite{Michel2018}.
%Such strongly nonlinear coherent structures being rich in harmonics (bound waves), a slowly random frequency modulation of such 1D anharmonic wave trains is known to be able to generate continuous frequency power-law spectrum with exponent between $-5$ and $-4$ \cite{Michel2018}. 
In a limit case, if these extreme events tend towards 1D singular coherent structures, their spectrum is predicted to scale as $\omega^{-4}$ \cite{Kuznetsov2004}. The wave field statistics is also reported revealing a heavy-tailed distribution that becomes independent of the distance after few nonlinear length scales of propagation. To sum up, most of these observations are compatible with the integrable turbulence theory for NLSE, but some deviations are also observed (power-law spectrum) related to the strong asymmetrical extreme events that exist in hydrodynamics. This hydrodynamics system is thus a good candidate to question the departure from the integrable turbulence theory in real systems (e.g. how the coherent structures close to integrability are deformed by bound waves). In the future, we plan to apply a local IST processing to identify and classify the different type of coherent structures in our time series, and their respective contributions to integrable turbulence~\cite{Randoux16b}.

%As in optical turbulence (NLS type), our hydrodynamic experiment appears to be a good candidate to show that coherent structures and large fluctuations are inevitable in order to balance fluxes and allow it to reach a statistically stationary state \cite{newell06}.

%although no prediction for the spectrum exists so far for the deep water regime (i.e. NLSE)
%and their contribution in the interplay with between wave turbulence and soliton turbulence.

%, involving dynamics and steeping of coherent pulses, probably occurs here. The low-frequency spectrum is ascribed to the presence of coherent structures such as envelope solitons. 

\begin{acknowledgments}
This work was supported by the French National Research Agency (ANR DYSTURB project No. ANR-17-CE30-0004). We thank S. Randoux and P. Suret for fruitful discussions.
\end{acknowledgments}

%Note that the maximum relative increase of $K-3$ is 100\% when changing the spectral width $\Delta f$ at fixed $\epsilon_\eta$, and is only 20\% when changing the steepness $\epsilon_\eta$ at fixed $\Delta f$ (not shown here). 

%\begin{figure}
%\includegraphics[width=7.5cm]{SkewnessTlTnl}
%\includegraphics[width=7cm]{KurtosisTlTnlNew}
%\caption{ Skewness (gauche) et kurtosis (droite) du signal de hauteur en fonction de la position adimensionnée, $x/L_{nl}$, de la sonde pour plusieurs valeurs de $T_l/T_{nl}$. $T_l/T_{nl}$=0.52 (en vert), 1.05 (en rouge), 1.65 (en bleu), 2.38 (en noir).}
%\label{StatSondes}
%\end{figure}

%  \begin{figure}
%   \centering
%   \includegraphics[width=9.5cm]{C8_Spatotemp06_02}
%  \caption{Signal de hauteur en différentes sondes représenté avec un décalage en hauteur proportionnel à l'abscisse de chaque sonde pour $T_l/T_{nl}=1.05$. La droite en pointillés rouges a pour pente la vitesse de groupe théorique d'ondes de Stokes. En noir est représenté l'ajustement avec la formule théorique d'un soliton. Même paramètres qu'en Fig.\ \ref{etavstime}.}
%  \label{beau2}
%  \end{figure} 

%%%%%%%%%%%%%%%%%%%%%%%%%%%%%%%%%%%%%%
%%%%%%%%%%%% REFERENCES %%%%%%%%%%%%%%%%%%
%%%%%%%%%%%%%%%%%%%%%%%%%%%%%%%%%%%%%%


\begin{thebibliography}{99}
%%%%%%%%%%%% Introduction
%%%%%%Wave turbulence
\bibitem{Falcon2010}E. Falcon, Laboratory experiments on wave turbulence, Discrete Cont. Dyn. B, {\bf 13} 819, (2010)	
\bibitem{Zakharovbook}V. E. Zakharov, V. L\'vov, and G. Falkovich, {\em Kolmogorov Spectra of Turbulence I: Wave Turbulence} (Springer-Verlag, Berlin, 1992)
\bibitem{Nazarenkobook}S. Nazarenko, {\em Wave Turbulence} (Springer, Berlin, 2011)
\bibitem{Newell2011}A. C. Newell and B. Rumpf, Wave turbulence, Annu. Rev. Fluid Mech. {\bf 43}, 59 (2011)
\bibitem{Hasselmann1962}K. Hasselmann, On the non-linear energy transfer in a gravity-wave spectrum Part 1. General theory, J. Fluid. Mech. {\bf 12}, 481 (1962)
\bibitem{Benney1967}D. J. Benney and A. C. Newell, The propagation of non-linear wave envelopes, J. Math. Phys. {\bf 46}, 363 (1967)
\bibitem{Zakharov1967}V. E. Zakharov and N. N. Filonenko, Energy spectrum for stochastic oscillations of the surface of liquid, Sov. Phys. Dokl. 11, 881-884 (1967) 
%V. E. Zakharov and N. N. Filonenko J. Appl. Mech. Tech. Phys. {\bf 8}, 37 (1967)
\bibitem{NazarenkoAdvance2013}V. Shrira and S. Nazarenko (Eds), {\em Advances in wave turbulence} Vol. {\bf 83} (World Scientific, Singapore, 2013)
\bibitem{Michel17}G. Michel, F. Pétrélis, and S. Fauve, Observation of thermal equilibrium in capillary wave turbulence, Phys. Rev. Lett. {\bf 118}, 144502 (2017)

%%%% Integrable turbulence or gas of soliton
\bibitem{Zakharov71}V. E. Zakharov, Kinetic equations for solitons, Sov. Phys. JETP {\bf 33}, 538 (1971) %KDV seulement
\bibitem{Zakharov09}V. E. Zakharov, Turbulence in integrable systems, Stud. Appl. Math. {\bf 122}, 219 (2009)
\bibitem{Randoux16}S. Randoux, P. Walczak, M. Onorato, and P. Suret, Nonlinear random optical waves: Integrable turbulence, rogue waves and intermittency, Physica D {\bf 333}, 323 (2016)
\bibitem{Kingsep73}A. S. Kingsep, L. I. Rudakov, and R. N. Sudan, Spectra of Strong Langmuir Turbulence, Phys. Rev. Lett. {\bf 31}, 1482 (1973)
\bibitem{Schwache97}A. Schwache and F. Mitschke, Properties of an optical soliton gas, Phys. Rev. E {\bf 55}, 7720 (1997) ; F. Mitsche, I. Halam, and A. Schwache, Soliton gas, Chaos Solitons Fractals {\bf 10}, 913 (1999)
\bibitem{Randoux14}S. Randoux, P. Walczack, M. Onorato, and P. Suret, Intermittency in integrable turbulence, Phys. Rev. Lett. {\bf 113}, 113902 (2014)
\bibitem{Toenger15}S. Toenger, T. Godin, C. Billet, F. Dias, M. Erkintalo, G. Genty, J.M. Dudley, Emergent rogue wave structures and statistics in spontaneous modulation instability, Sci. Rep. {\bf 5}, 10380 (2015)
\bibitem{Walczak15}P. Walczak, S. Randoux, and P. Suret, Optical rogue waves in integrable turbulence, Phys. Rev. Lett. {\bf 114}, 143903 (2015)
\bibitem{Suret16}P. Suret, R. El Koussaifi, A. Tikan, C. Evain, S. Randoux, C. Szwaj, and S. Bielawski, Single-shot observation of optical rogue waves in integrable turbulence using time microscopy, Nature Com. {\bf 7}, 13136 (2016)
%non integrable turbulence (soliton turbulence)
%\bibitem{Zakharov88}V. E. Zakharov, A. N. Pushkarev, V. F. Shvets, V. V. Yankov, Soliton turbulence, JETP Lett. {\bf 48}, 79 (1988) ; A. I. Dyachenko, V. E. Zakharov, A. N. Pushkarev, V. F. Shvets, V. V. Yankov, Soliton turbulence in nonintegrable wave systems, JETP Lett. {\bf 96} 2026 (1989)


\bibitem{Chabchoub2013} A. Chabchoub, O. Kimmoun, H. Branger, N. Hoffmann, D. Proment, M. Onorato and N. Akhmediev, Experimental Observation of Dark Solitons on the Surface of Water,  Phys. Rev. Lett. \textbf{110}, 124101 (2013).


\bibitem{Osborne93}A. R. Osborne, Behavior of solitons in random-function solutions of the periodic Korteweg-de Vries equation, Phys. Rev. Lett. {\bf 71}, 3115 (1993)
\bibitem{Osborne91}A. R. Osborne, E. Segre, G. Boffetta, and L. Cavaleri, Soliton basis states in shallow-water ocean surface waves, Phys. Rev. Lett. {\bf 67}, 592 (1991)
\bibitem{Costa14}A. Costa, A. R. Osborne, D. T. Resio, S. Alessio, E. Chrivì, E. Saggese, K. Bellomo, and C. E. Long, Soliton Turbulence in Shallow Water Ocean Surface Waves, Phys. Rev. Lett. {\bf 113}, 108501 (2014)
\bibitem{Perrard15}S. Perrard, L. Deike, C. Duch\^ene, and C. T. Pham, Capillary solitons on a levitated medium, Phys. Rev. E {\bf 92}, 011002(R) (2015)
\bibitem{Hassaini17}R. Hassaini, and N. Mordant, Transition from weak wave turbulence to soliton gas, Phys. Rev. Fluids {\bf 2}, 094803 (2017)
%\bibitem{Newell12}A. C. Newell, B. Rumpf, and V. E. Zakharov, Spontaneous Breaking of the Spatial Homogeneity Symmetry in Wave Turbulence, Phys. Rev. Lett.  {\bf 108}, 194502 (2012)
%\bibitem{Rumpf09}B. Rumpf, A. C. Newell, and V. E. Zakharov, Turbulent Transfer of Energy by Radiating Pulses, Phys. Rev. Lett. {\bf 103}, 074502 (2009)
\bibitem{Onorato04}M. Onorato, A. R. Osborne, M. Serio, L. Cavaleri, C. Brandini and C. T. Stansberg, Observation of strongly non-Gaussian statistics for random sea surface gravity waves in wave flume experiments, Phys. Rev. E {\bf 70}, 067302 (2004) ; M. Onorato, A. Osborne, M. Serio, L. Cavaleri, Modulational instability and non-gaussian statistics in experimental random water-wave trains, Phys. Fluids {\bf 17}, 078101 (2005)
\bibitem{Shemer09}L. Shemer and A. Sergeeva, An experimental study of spatial evolution of statistical parameters
in a unidirectional narrow-banded random wavefield, J. Geophys. Res. {\bf 114}, C01015 (2009);
\bibitem{ShemerPoF10}L. Shemer, A. Sergeeva, and A. Slunyaev, Applicability of envelope model equations for simulation of narrow-spectrum unidirectional random wave field evolution: Experimental validation, Phys. Fluids, {\bf 22} 016601 (2010)
\bibitem{ShemerJGR10}L. Shemer, A. Sergeeva, and D. Liberzon, Effect of the initial spectrum on the spatial evolution of statistics of unidirectional nonlinear random waves, J. Geophys. Res. {\bf 115}, C12039 (2010); 
 \bibitem{ElKoussaifiPRE18}R. El Koussaifi, A. Tikan, A. Toffoli, S. Randoux, P. Suret and M. Onorato, Spontaneous emergence of rogue waves in partially coherent waves: a quantitative experimental comparison between hydrodynamics and optics, Phys. Rev. E {\bf 97}, 012208 (2018)
\bibitem{JanssenJPO03}P. A. E. M Janssen, Nonlinear Four-Wave Interactions and Freak Waves, J. Phys. Oceano. {\bf 33}, 863 (2003)
\bibitem{Onorato01}M. Onorato, A.R. Osborne, M. Serio, S. Bertone, Freak Waves in Random Oceanic Sea States, Phys. Rev. Lett. {\bf 86}, 5831 (2001) ; M. Onorato, A.R. Osborne, M. Serio, Extreme wave events in directional, random oceanic sea states, Phys. Fluids {\bf 14}, L25 (2002)
\bibitem{Dysthe03}K. B. Dysthe, K. Trulsen, H.E. Krogstad, H. Socquet-Juglard, Evolution of a narrow-band spectrum of random surface gravity waves, J. Fluid Mech. {\bf 478}, 1 (2003)
\bibitem{Slunyaev06}A. Slunyaev, Nonlinear analysis and simulations of measured freak wave time series, Eur. J. Mech. B/Fluids {\bf 25}, 621 (2006)
\bibitem{SotoCrespo16}J. M. Soto-Crespo, N. Devine, and N. Akhmediev, Integrable turbulence and rogue waves: Breathers or solitons?, Phys. Rev. Lett. {\bf 116}, 103901 (2016)
%Used the Inverse Scattering Technique (IST) limited only to the integrable models
%G. El, A. Kamchatnov, Kinetic equation for a dense soliton gas, Phys. Rev. Lett. 95(20) (2005) 204101.
%G.A. El, A.M. Kamchatnov, M.V. Pavlov, S.A. Zykov, Kinetic equation for a soliton gas and its hydrodynamic reductions, J.Nonlinear Sci. 21(2) (2011) 151–191.
\bibitem{Osborne05}A. R. Osborne, M. Onorato, M. Serio, Nonlinear Fourier analysis of deep-water, random surface waves: Theoretical formulation and experimental observations of rogue waves, in Proc. of Hawaiian Winter Workshop: Rogue Waves, 2005.
\bibitem{Islas05}A.L. Islas, C.M. Schober, Predicting rogue waves in random oceanic sea states, Phys. Fluids {\bf 17} 031701 (2005)
\bibitem{Whitham}G. B. Whitham, {\em Linear and Nonlinear Waves}, John Wiley \& Sons Inc. (1974)

%%%%%%%%%%%%%
%%%%%%%%%%%%%%
\bibitem{Remoissenet99}M. Remoissenet, {\it Waves Called Solitons: Concepts and Experiments}, 3rd. Ed, (Berlin, Springer-Verlag, 1999)
\bibitem{Benney67}D. J. Benney and A. C. Newell, The propagation of nonlinear waves envelope, J. Math. Phys. {\bf 46}, 133 (1967)
\bibitem{Zakharov68}V. E. Zakharov, Stability of periodic waves of finite amplitude on the surface of a deep fluid, J. Appl. Mech. Tech. Phys. {\bf 9}, 86 (1968)
\bibitem{ZakharovShabat72}V. E. Zakharov and A. B. Shabat, Exact theory of two-dimensional self-focusing and one dimensional self-modulation of waves in nonlinear media, Sov. Phys. JETP {\bf 34}, 62 (1972)
\bibitem{Yuen75}H. C. Yuen and B. C. Lake, Nonlinear deep water waves: Theory and experiment. Phys. Fluids {\bf 18}, 956 (1975)
\bibitem{Yagi76}T. Yagi and A. Noguchi, Experimental studies on modulation instability by using nonlinear transmission lines, Elec. and Commun. Japan {\bf 59A}, 1 (1976) ; Gyromagnetic nonlinear element and its application as a pulse-shaping transmission line, Electron. Let. {\bf 13}, 683 (1977)

\bibitem{ChabchoubPOF2016}A. Chabchoub and R. H. J. Grimshaw, The hydrodynamic Nonlinear Schrödinger equation: Space and time, Phys. Fluids {\bf 1}, 23 (2016)
\bibitem{Peregrine83}D. H. Peregrine, Water waves, nonlinear Schrödinger equations and their solutions, J. Aust. Math. Soc. Series B Appl. Math. {\bf 25}, 16 (1983)
\bibitem{Kuznetsov}E. Kuznetsov, Solitons in a parametrically unstable plasma. Akademiia Nauk SSSR Doklady {\bf 236}, 575 (1977)
\bibitem{Ma}Y. C. Ma, The perturbed plane-wave solutions of the cubic Schrödinger equation, Stud. Appl. Math. {\bf 60}, 43 (1979)
\bibitem{Akmediev1985}N. Akhmediev, V. M. Eleonskii, N. E. Kulagin, Generation of periodic trains of picosecond pulses in an
optical fiber: Exact solutions, Sov. Phys. JETP {\bf 62}, 894 (1985)
\bibitem{Akmediev1987}N. Akhmediev, V. M. Eleonskii, N. E. Kulagin, Exact solutions of the first order of nonlinear Schrödinger
equation, Theor. Math. Phys. (USSR) {\bf 72}, 809 (1987)
\bibitem{Chabchoub11}A. Chabchoub, N. P. Hoffmann and N. Akmediev, Rogue wave observation in a water wave tank, Phys. Rev. Lett. {\bf 106}, 204502 (2011)
\bibitem{Slunyaev13}A. Slunyaev, G. F. Clauss, M. Klein, and M. Onorato, Simulations and experiments of short intense envelope solitons of surface water waves, Phys. Fluids {\bf 25}, 067105 (2013); A. Slunyaev, M. Klein, and G. F. Clauss, Laboratory and numerical study of intense envelope solitons of water waves: Generation, reflection from a wall, and collisions, Phys. Fluids {\bf 29}, 047103 (2017)
\bibitem{KiblerNature10}B. Kibler, J. Fatome, C. Finot, G. Millot, F. Dias, G. Genty, N. Akhmediev, and J. M. Dudley, The Peregrine soliton in nonlinear fibre optics, Nat. Phys {\bf 6}, 790 (2010)
\bibitem{BailungPRL11}H. Bailung, S. K. Sharma, and Y. Nakamura, Observation of Peregrine Solitons in a Multicomponent Plasma with Negative Ions, Phys. Rev. Lett. {\bf 107}, 255005 (2011)

\bibitem{Annexe0}Since $\Omega/ K=c_g=\omega_0/(2k_0)$, one uses $\Omega /\omega_0=K/(2k_0)$ in the condition for modulation.
\bibitem{Lighthill65}M. J. Lighthill, Contribution to the theory of waves in nonlinear dispersive systems, J. Inst. Math. Appl. 1, {\bf 269} (1965)

\bibitem{BenjaminFeir67}T. B. Benjamin and J. E. Feir, The disintegration of wave trains on deep water. Part 1 Theory, J. Fluid Mech {\bf 27}, 417 (1967)
\bibitem{Lake77}B. M. Lake, H. C. Yuen, H. Rungaldier and W. E. Ferguson, Nonlinear deep water waves:theory and experiment. 2. Evolution of a continuous wave train. J. Fluid. Mech. {\bf 83}, 49 (1977)
\bibitem{LonguetHiggins80}M. S. Longuet-Higgins, Modulation of the amplitude of steep wind waves, J. Fluid Mech. {\bf 99}, 705 (1980)
\bibitem{Melville82}W. K. Melville, The instability and breaking of deep-water waves. J. Fluid Mech. {\bf 115}, 165 (1982)
\bibitem{Su82}M. Y. Su, Evolution of groups of gravity waves with moderate to high steepness, Phys. Fluids {\bf 25}, 2167 (1982)
\bibitem{Annexe}For a Gaussian pulse, $|A(\xi,0)|=A_0\exp{(-\xi^2/2L_0^2)}$ with $L_0$ its half-width (at an amplitude of $A_0/\sqrt{e}$), the spatial dispersion of the Gaussian wave packet at time $T$, governed by Eq.\ (\ref{NLS}) with $Q=0$, reads $L(T)=L_0\sqrt{1+(T/T_{lin})^2}$ with the typical dispersion time $T_{lin}=L_0^2/(2P)$ \cite{Remoissenet99}. Although its broadening, the pulse energy $\int_{-\infty}^{+\infty} |A(\xi,T)|^2d\xi=\int_{-\infty}^{+\infty} |A(\xi,0)|^2d\xi$ is conserved during the propagation as well as its Gaussian shape. 
%\bibitem{Annexe2}For a Gaussian pulse, the full width at half-height is $L_{fw}=2\sqrt{\ln 2}L_{{\rm sol}}$. For a $\sech$ pulse $A(X,0)=A_0\sech(X/L_{{\rm sol}})$, the full width at half-height is $L_{fw}=L_{{\rm sol}}/\sech(1/2)$ \cite{Remoissenet99}.

\bibitem{Chabchoub15}A. Chabchoub, B. Kibler, C. Finot, G. Millot, M. Onorato, J. Dudley, and A. Babanin, The nonlinear Schrödinger equation and the propagation of weakly nonlinear waves in optical fibres and on the water surface, Annals of Physics {\bf 361}, 490 (2015)
\bibitem{Annexe3}In the reference frame moving at the group velocity $c_g$, on a has $\Delta \omega=c_g\Delta k$, with $c_g=\omega_0/(2k_0)$ leading to $\Delta \omega/\omega_0=\Delta k/(2k_0)$, and the expression of $BFI_{\omega}$ from $BFI_k$.
\bibitem{BonnefoyJFM16}F. Bonnefoy, F. Haudin, G. Michel, B. Semin, T. Humbert, S. Auma\^{\i}tre, M. Berhanu, and E. Falcon, Observation of resonant interactions among surface gravity waves, J. Fluid Mech. {\bf 805}, R3 (2016)
\bibitem{Annexe4}To be consistent with the estimate of $T_{lin}$, the full width at half maximum of the wave packet, $L$, is defined in Eq.\ (\ref{NLS}) by $2\beta/L^2 \equiv \partial^2 /\partial x^2$, $\beta$ depending on its shape (e.g. $\beta=2\arcsech{1/2}$ for a sech-pulse, $\beta=2\sqrt{2\ln{2}}$ for Gaussian). If $L$ is the half width of a Gaussian at a $1/\sqrt{e}$ amplitude, then $\beta=1$.

\bibitem{TikanPRL17}A. Tikan, et al., Universality of the Peregrine soliton in the focusing dynamics of the cubic nonlinear Schr{\"o}dinger equation, Phys. Rev. Lett. {\bf 119}, 033901 (2017) ; A. Tikan, S. Bielawski, C. Szwaj, S. Randoux, and P. Suret, Single-shot measurement of phase and amplitude by using a heterodyne time-lens system and ultrafast digital time-holography, Nat. Photon. {\bf 12}, 228 (2018)
\bibitem{ShemerPoF13}L. Shemer and L. Alperovich, Peregrine breather revisited, Phys. Fluids {\bf 25}, 051701 (2013)
\bibitem{DongPRF18}G. Dong, B. Liao, Y. Ma, and M. Perlin, Experimental investigation of Peregrine breather of gravity waves on finite water depth, Phys. Rev. Fluids {\bf 3}, 064801 (2018)
\bibitem{ChabchoubPoF2013}A. Chabchoub, N. Hoffmann, H. Branger, C. Kharif, and N. Akhmediev, Experiments on wind-perturbed rogue wave hydrodynamics using the Peregrine breather model, Phys. Fluids {\bf 25}, 101704 (2013)
\bibitem{ChabchoubProceed17}A. Chabchoub, G. Genty, J. M. Dudley, B. Kibler, and T. Waseda, Experiments on spontaneous modulation instability in hydrodynamics, Proceedings ISOPE, ISOPE-I-17-582, pp. 420 -- 424 (2017)
\bibitem{ChabchoubPRL16}A. Chabchoub, Tracking breather dynamics in irregular sea state conditions, Phys. Rev. Lett. {\bf 117}, 144103 (2016)
\bibitem{RandouxPreprint18}S. Randoux, P. Suret, A. Chabchoub, B. Kibler, and G. El, Nonlinear spectral analysis of Peregrine solitons observed in optics and in hydrodynamic experiments, arXiv:1806.10785 (2018)

\bibitem{Randoux16b}S. Randoux, P. Suret and G. El, Inverse scattering transform analysis of rogue waves using local periodization procedure, Sci. Rep. {\bf 6}, 29238 (2016)

\bibitem{InfeldBook}E. Infeld and G. Rowlands, {\em Nonlinear Waves, Solitons and Chaos} (Cambridge Univ. Press, New York, 2nd ed., 2002)
\bibitem{AkhmedievPRE09}N. Akhmediev, A. Ankiewicz, and J. M. Soto-Crespo, Rogue waves and rational solutions of the nonlinear Schr{\"o}dinger equation, Phys. Rev. E {\bf 80}, 026601 (2009)

\bibitem{Duncan99}J. H. Duncan, H. Qiao, V. Philomin, and A. Wenz, Gentle spilling breakers: crest profile evolution, J. Fluid Mech. {\bf 379}, 191 (1999) and references therein
\bibitem{Falcon10}E. Falcon, S. G. Roux and C. Laroche, On the origin of intermittency in wave turbulence, EPL {\bf 90}, 34005 (2010)
\bibitem{Zakharov04}V. Zakharov, F. Dias, and A. Pushkarev, One-dimensional wave turbulence, Phys. Rep. {\bf 398}, 1 (2004)
\bibitem{Dyachenko95}A. Dyachenko, Y. Lvov, V. Zakharov, Five-wave interaction on the surface of deep fluid, Physica D {\bf 87}, 233 (1995)
\bibitem{Connaughton03}C. Connaughton, S. Nazarenko, A. C. Newell, Dimensional analysis and weak turbulence, Physica D {\bf 184}, 86 (2003)
\bibitem{Michel2018}G. Michel, B. Semin, A. Cazaubiel, F. Haudin, T. Humbert, S. Lepot, F. Bonnefoy, M. Berhanu, and E. Falcon, Self-similar gravity wave spectra resulting from the modulation of bound waves, Phys. Rev. Fluids {\bf 3}, 054801 (2018)
\bibitem{Kuznetsov2004}E.A. Kuznetsov, Turbulence spectra generated by singularities, JETP Lett. {\bf 80}, 83 (2004); S. Nazarenko and S. Lukaschuk and S. McLelland and P. Denissenko, Statistics of surface gravity wave turbulence in the space and time domains, J. Fluids Mech. {\bf 642}, 395 (2010)

\bibitem{Tayfun80}M. A. Tayfun, Narrow-band nonlinear sea waves, J. Geophys. Res. {\bf 85}, 1548 (1980)
\bibitem{Falcon07}E. Falcon, C. Laroche, and S. Fauve, Observation of gravity-capillary wave turbulence, Phys. Rev. Lett. {\bf 98}, 094503 (2007); E. Falcon and C. Laroche, Observation of depth-induced properties in wave turbulence, EPL {\bf 95}, 34003 (2011); B. Issenmann and E. Falcon, Gravity wave turbulence revealed by horizontal vibrations of the container, Phys. Rev. E {\bf 87}, 011001(R) (2013)
\bibitem{Forristall00}G. Z. Forristall, Wave crest distributions: Observations and second-order theory, J. Phys. Oceano. {\bf }, 1931 (2000)
\bibitem{Evans00}M. Evans, N. Hastings, and B. Peacock, {\em Statistical distributions}, (3rd Ed., Wiley \& sons, New York, 2000)

\bibitem{Falkovitch06}G. Falkovitch, Lecture course for the Warwick Summer School, July, 2006
%\bibitem{Madja97}A. J. Majda, D.W. McLaughlin, and E. G. Tabak, A one-dimensional model for dispersive wave turbulence, J. Nonlinear Sci. {\bf 7}, 9 (1997).
%\bibitem{Slunyaev10}A. Slunyaev and A. Starobor, AKNS eigenvalue spectrum for densely spaced envelope solitary waves, Geophysical Research Abstracts {\bf 12}, EGU2010-1731 (2010)
%\bibitem{newell06}A. C. Newell, Lecture in Rencontre du Non Linéaire, Paris 2006

 \end{thebibliography}
\end{document}